\documentclass[english,12pt]{article}

\usepackage{geometry}
\usepackage{xcolor}
\usepackage{array}
\usepackage{calc}
\usepackage{amsmath}
\usepackage{amssymb}
\usepackage{graphicx}
\usepackage{setspace}
\usepackage{natbib}
\usepackage{soul}
\usepackage{nameref}
\usepackage{float}
\date{}
\usepackage{babel}
\usepackage{hyperref}
\usepackage{hhline}

\usepackage{patchcmd}
\patchcommand\align{\hrule height 0pt depth 0pt width 0pt}{}
\patchcommand\multline{\hrule height 0pt depth 0pt width 0pt}{}
\usepackage{caption}

\usepackage{float}
\floatstyle{ruled}
\newfloat{floatboxX}{t}{lob}
\floatname{floatboxX}{Box}
\newenvironment{floatbox}{\begin{floatboxX}\sf\small\setlength{\parindent}{15pt}
}{\end{floatboxX}}

\usepackage{multicol}

\usepackage{tikz}
\usetikzlibrary{arrows}
\usetikzlibrary{decorations.pathmorphing,shapes}
\usetikzlibrary{snakes}
\newcounter{sarrow}

\makeatletter

\newcommand{\beginappendix}{%
    \setcounter{table}{0}
    \renewcommand{\thetable}{\thesection.\arabic{table}}%
    \setcounter{figure}{0}
    \renewcommand{\thefigure}{\thesection.\arabic{figure}}%
    \setcounter{equation}{0}
    \renewcommand{\theequation}{\thesection.\arabic{equation}}%
}
\makeatother

\usepackage{xr}
\makeatletter
\newcommand*{\addFileDependency}[1]{
  \typeout{(#1)}
  \@addtofilelist{#1}
  \IfFileExists{#1}{}{\typeout{No file #1.}}
}
\makeatother

\DeclareMathOperator{\erf}{erf}

\begin{document}

\title{A metric for tradable biodiversity credits linked to the Living Planet Index and global species conservation}

\author{Axel G.\ Rossberg$^{1}$, Jacob D.\ O'Sullivan$^{1}$,
  Svetlana Malysheva$^{3}$\\ and Nadav M.\ Shnerb$^{4}$}

\date{\it \today}
\maketitle
\vspace{-3em}
\rule[0.5ex]{1\columnwidth}{1pt}
\small{
$^{1}$School of Biological and Chemical Sciences, Queen Mary University
of London, Mile End Road, London, E1 4NS, United Kingdom\\
$^{2}$Department of Environmental Science, Faculty of Sciences, Radboud University, Nijmegen, The Netherlands\\
$^{3}$School of Mathematical Sciences, Queen Mary University
of London, Mile End Road, London, E1 4NS, United Kingdom\\
$^{4}$Department of Physics, Bar-Ilan University, Ramat-Gan IL52900, Israel
\bigskip{}

\noindent\textbf{Corresponding author:} Axel G. Rossberg, 
School of Biological and Behavioural Sciences, Queen Mary University of London, Mile End Road,
London E1 4NS, United Kingdom. (\href{mailto:a.rossberg@qmul.ac.uk}{a.rossberg@qmul.ac.uk}, +447551396243)


}
\newpage



\begin{abstract} 
  The difficulty of identifying appropriate metrics for the impact organisations have on biodiversity remains a major barrier to the inclusion of biodiversity considerations in environmentally responsible investment.
  We derive and analyse a simple metric: the sum of the proportional changes in the global abundances of species caused by an organisation, with a regularisation to cover the case of species close to extinction.
  We argue mathematically that this metric quantifies changes in the mean long-term global survival probability of species.
  The metric thus supports the objective ``to significantly reduce extinction risk'' of the new Global Biodiversity Framework and allows organisations to set themselves a corresponding science-based target.
  It also quantifies impact on a widely cited global biodiversity indicator, the Living Planet Index, for which we propose an improved formula that naturally resolves the known problem of singularities caused by extinctions.
  We show that in a perfect market trade in biodiversity credits quantified by our metric would lead to near-optimal allocation of resources to species conservation.
  We further show that metric values are quantitatively convertible to several other metrics and indices already in use.
  Barriers to adoption are therefore low.
  Used in conjunction with metrics addressing ecosystem extent and integrity, potential areas of application include biodiversity related financial disclosures and voluntary or legislated no net biodiversity loss policies.
\end{abstract}



\section{Introduction}

The rapid recent growth of markets for responsible investments considering Environmental, Social and Governance (ESG) concerns \citep{Diab21:_ESGAssets}, and the growing attention biodiversity receives in this context \citep{Addison19:_UsingConservation, Nauman20:_ESGInvestors}, highlight the need for metrics of biodiversity impacts appropriate for use by businesses and  financial institutions \citep{Addison19:_UsingConservation, Smith20:_BiodiversityMeans, TNFD21:_ProposedTechnical}.

\begin{table}
  \caption{\label{tab:desiderata}Properties of biodiversity metrics considered important in the business context, how Biodiversity Impacts Credits (BICs) address these, and where we demonstrate them.}
  
\center
\begin{tabular}{|p{0.3\linewidth}|p{0.53\linewidth}|c|}
  \hline
   Needs we address& How we achieve this& Section\\
  \hhline{|===|} 
  Scientific robustness  \citep{TNFD22:_TNFDNature-Related}&  BICs are  mathematically linked to long-term species extinction risk. &\ref{sec:theory}\\
  \hline
   Assessment of both negative and positive impacts \citep{Bor18:_PositiveImpact}& BICs naturally quantify both positive and negative impacts.&\ref{sec:theory}\\
  \hline
  Assessment of biodiversity risks and opportunities  \citep{TNFD22:_TNFDNature-Related}& The close link of BICs to  long-term extinction risk facilitates their use in probabilistic risk assessments. By quantifying system state (rather than projections), BICs incentivise innovation where such opportunities arise.&\ref{sec:theory}, \ref{sec:discussion}\\
  \hline
  Comparability across sectors and providing insights to inform corporate and financial institution decision making, including at aggregated portfolio levels for financial institution  \citep{TNFD22:_TNFDNature-Related}&BICs measure biodiversity impact in a single currency based on a simple formula and can be aggregated from site to portfolio level. Decision making based on BICs pricing aligns with conservation objective.&\ref{sec:theory},  \ref{sec:discussion}\\
  \hline
  Use at scale and at different levels  \citep{TNFD22:_TNFDNature-Related}&  BICs can be estimated at site, product, or corporate level using existing metrics and tools.& \ref{sec:calculation}\\
  \hline
  Use for corporate target setting  \citep{TNFD22:_TNFDNature-Related}& Corporations can use BICs to set and demonstrably achieve net-zero or net-positive biodiversity  targets.&\ref{sec:theory}\\
  \hline
  Alignment with broader national and global public policy goals for reversing nature loss  \citep{TNFD22:_TNFDNature-Related}&  BICs are aligned with the global policy objective of reducing extinction risk and with widely used geometric mean abundance indices such as the global Living Planet Index (LPI)  \citep{WWF22:_LivingPlanet} or UK's Wild-Bird Indicator  \citep{Defra21:_WildBird}.&\ref{sec:materials-methods}, \ref{sec:theory}\\
  \hline
  Suitability for biodiversity credit markets  \citep{WorldEconomicForum22:_BiodiversityCredits}& BICs are additive, so total BICs do not change through paper transactions. Market forces generated by BICs align with global conservation objectives.&\ref{sec:theory}\\
  \hline
\end{tabular}
\end{table}

Requirements on such metrics include most of those identified in the ecosystem management and scientific context \citep{Rice05:_FrameworkSelecting}, listed, e.g., in the author guidelines of \emph{Ecological Indicators} (\url{https://www.sciencedirect.com/journal/ecological-indicators}), such as simplicity, relevance, scientific justifiability, cost-efficiency, and reliable quantification.
The business community, however, emphasises other needs, some listed in the first column of Table~\ref{tab:desiderata}.
These derive from aspects of the business world not encountered as such in traditional conservation, including continuous innovation, risk taking, hierarchical structures of ownership and accountability, transactions under private law, and decision-making in a fluid, complex socioeconomic context.
%
Important in the business context is also to understand market responses to quantification of biodiversity impacts by a particular metric---in particular whether this might create unintentional, perverse incentives \citep{Ran13:_PerverseIncentive}.


The 23 Targets of the \cite{Kunming-MontrealGlobalBiodiversityFramework22:_} provide orientation to business on ecological and societal biodiversity targets that metrics should consider.
The present study focuses on metrics related to the aim of Target 4 ``to significantly reduce extinction risk''.
We propose two closely related metrics, which we call \emph{Biodiversity Impact Credits} (BIC) and \emph{Biodiversity Stewardship Credits} (BSC), that quantify impacts on mean long-term species survival probability.
For a good representation of business impacts on biodiversity, these will need to be combined with metrics addressing other objectives, such as Target~2, aiming to put ``at least 30 per cent of areas of degraded terrestrial, inland water, and coastal and marine ecosystems [...] under effective restoration, in order to enhance biodiversity and ecosystem functions and services, ecological integrity and connectivity.''
Metrics suitable for Target~2 might include simple metrics of spatial extent or, to include representation of ecosystem function and ecological integrity \citep{Mace05:_IndexIntactness}, the Mean Species Abundances metric (MSA, \citealt{Alkemade09:_GLOBIO3Framework}), given by the arithmetic mean local proportional population decline of species relative to the natural baseline.
Alternatively, one can use composite metrics tailored to regional particularities and needs such as the `ecosystem credits' computed using the Biodiversity Assessment Method of New South Wales \citep{NSWDepartmentofPlanningandEnvironment20:_BiodiversityAssessment} or Natural England's Biodiversity Metric \citep{NaturalEngland21:_BiodiversityMetric}.

Even in the narrower context of metrics addressing species conservation, BICs and BSCs do not stand alone \citep{Williams96:_ComparisonRichness, Muller-Wenk98:_LandUse, Butchart07:_ImprovementsRed, McRae17:_DiversityweightedLiving,Mair21:_MetricSpatially}.
What distinguishes our proposal for BICs and BSCs from previous proposals is that we shall derive these metrics mathematically from a mathematical model for extinction risk.
Furthermore, we demonstrate below how BICs and BSCs are aligned through approximate mathematical relations to several widely used and cited metrics addressing species conservation. Specifically, i) BICs quantify impacts on the \emph{Living Planet Index} (LPI) \citep{WWF22:_LivingPlanet,McRae17:_DiversityweightedLiving}, one of the most widely cited global biodiversity indicators currently in use (4,710 publications identified by Google Scholar per July 2022); ii) BSCs are strongly correlated with the \emph{Species Threat Abatement and Recovery} (STAR, 21 publications) metric \citep{Mair21:_MetricSpatially}, a spatially resolved variant of the IUCN's Red List Index (\citealt{Butchart07:_ImprovementsRed,Butchart10:_GlobalBiodiversity}; 2,490 publications); iii) BICs are approximated by life-cycle impact assessments scores based on the \emph{Potentially Disappeared Fraction} (PDF) of species \citep[][ 1,680 publications]{Muller-Wenk98:_LandUse, Goedkoop00:_Eco-Indicator99}; and iv) BSCs can be approximated by \emph{Range-Size Rarity} (\citealt{Williams96:_ComparisonRichness}; 885 publications, including name variants), a metric used in conservation ecology to identify sites for protected areas.
Through this quantitative unification of previously separated approaches we aim to add credence and scientific support to all metrics involved.
BICs contribute mathematical rigour to this group of metrics, while others contribute, e.g., existing demonstrations of utility, cost-efficiency, and reliable quantification.

The plan for this paper is as follows: In Section~\ref{sec:materials-methods} we will first establish the theoretical foundation of BICs in a model for species extinction risk and demonstrate the close relation of this model to the LPI\@.
In doing so, we obtain a solution to a long-standing conceptual problem arising with use and interpretation of the LPI when populations of species entering the LPI approach zero \citep{Collen09:_MonitoringChange, McRae17:_DiversityweightedLiving, Leung20:_ClusteredCatastrophic}.
We also briefly describe a model that we use to test and evaluate our metrics.
In Section~\ref{sec:theory} we will then derive the BIC metric as a measure of local impact on global species extinction risk.
We study the incentive structure generated by BICs in a perfect market, to make sure BICs do not generate perverse incentives.
Then BSCs are derived as a special case of BICs.
Finally, we use simulations to address questions of metric behaviour in light of metacommunity dynamics.
In Section~\ref{sec:calculation} we derive and describe several methods of metric computation. This is facilitated by the close mathematical relations between BIC, BSC and other metrics, which we demonstrate.
Discussing the results in Section~\ref{sec:discussion}, we emphasise use of BSCs and BICs in the business context, building on results from previous sections (Table~\ref{tab:desiderata}).

\section{Conceptual background, materials and methods}
\label{sec:materials-methods}

\subsection{Estimating the long-term extinction risk of species}
\label{sec:estimation-long-term}

We consider a simple mathematical model that allows us to analytically link the populations sizes of species to their long-term survival.
For a given taxonomic or functional group of species (below `group of species' or similar), denote for each species $i$ in that group by $N_i$ the global population size of that species.
Population sizes can be measured in a variety of ways, such as in terms of the number of mature individuals or population biomass, in some cases even by the number of colonies, whichever approximates total reproductive value \citep{rossberg11:_qna} well.

It has been shown on long-term time series data, e.g., for trees \citep{Kalyuzhny14:_NicheNeutrality, Kessler15:_NeutralDynamics}, fish \citep{Kessler15:_NeutralDynamics}, herbaceous plants \citep{Kessler15:_NeutralDynamics}, and birds \citep{Kalyuzhny14:_NicheNeutrality} that the populations of most species exhibit random walks on the $\log(N)$ axis with mean-squared increments that are largely independent of $N$, a phenomenon known as `environmental stochasticity' (caused by variability in both the abiotic environment and the abundances of co-occurring species).
For a given population, we denote the mean square of this increment during a time interval of length $\Delta t$, divided by $\Delta t$, by $v_{\mathrm{e}}$.
Since the same studies reveal little if any detectable drift towards smaller population sizes or any preferred value, we disregard such drift here.

Next to environmental stochasticity, the discrete nature of life-history events (e.g.\ birth, germination, death etc\@.)
generates additional variability in population sizes, known as `demographic stochasticity', for which mean-squared increments in $N$ over a given time interval $\Delta t$ are proportional to $N$.
We denote the corresponding proportionality constant, divided by $\Delta t$, by $v_{\mathrm{d}}$.
As a driver of fluctuations in species abundance, demographic stochasticity is widely understood to be negligible compared to environmental stochasticity, except for small populations \citep{Melbourne12:_StochasticityDemographic}.
The ratio $N^*=v_{\mathrm{d}}/v_{\mathrm{e}}$ specifies the population size below which demographic stochasticity dominates over environmental stochasticity.


\begin{floatbox}
  \caption{A simple model for species survival probability\label{box:regularisation}}
  \scriptsize 
  \begin{multicols}{2}
    We model the change in the population size of a species over a time interval $\Delta t$ as
    \begin{align}
      \label{eq:drift-model}
      N(t+\Delta t)=\exp\left[\xi(t)\sqrt{v_{\mathrm{e}}   \Delta t} \right]\, N(t) +  \xi'(t) \sqrt{v_{\mathrm{d}}\Delta t} N(t),
    \end{align}
    where $\xi(t)$ and $\xi'(t)$ denote independent standard normal random numbers. Parameters $v_{\mathrm{e}}$ and $v_{\mathrm{d}}$ represent the strengths of environmental and demographic stochasticity, respectively.
    Formally taking the limit $\Delta t \to 0$, standard procedures \citep{gardiner90:_handb_stoch_method} lead to an approximation of this process by the Ito stochastic differential equation
    \begin{align}
      \label{eq:ito-N}
      d N = \frac{v_{\mathrm{e}}}{2} N dt + \sqrt{v_{\mathrm{d}}N+ v_{\mathrm{e}}N^2} dW_t,
    \end{align}
    where $W_t$ represents a Wiener process (Brownian motion).
    The first term on the right-hand-side describes drift to larger values. It goes back to the fact that the expectation value of the log-normal distribution $\exp[\xi(t)\sqrt{v_{\mathrm{e}} \Delta t} ]$ in Eq.~\eqref{eq:drift-model} is $\exp(v_{\mathrm{e}} \Delta t/2)$.
    If one were to formulate this process in terms of $\log(N)$ rather than $N$, this term would disappear.
    However, demographic stochasticity would then generate another drift term instead.
    To eliminate drift altogether, we express population sizes in terms of $x=\log(1+N/N^*)$, where $N^*=v_{\mathrm{d}}/v_{\mathrm{e}}$ is the population size at which environmental and demographic stochasticity have the same strength (see Eq.~\eqref{eq:ito-N}).
    Note that $x$ becomes zero when a species goes extinct ($N=0$).
    Applying Ito's formula \citep{gardiner90:_handb_stoch_method}, this change of variables simplifies Eq.~\eqref{eq:ito-N} to
    \begin{align}
      \label{eq:ito-x}
      d x = \sqrt{(1-e^{-x}) v_{\mathrm{e}}} d W_t.
    \end{align}
    That is, $x$ performs a Brownian motion, represented by $dx = \sqrt{v_{\mathrm{e}}} dW_t$, except when $x$ is of the order of one or smaller.

    For $x$ approaching zero from above, the factor $1-e^{-x}$ reduces the magnitude of fluctuations in $x$, slowing down the random walk.
    As a result, $x$ can get trapped in the region of low $x$, and the vicinity of $0$ acts similar to an absorbing boundary  \citep{Pechenik99:_InterfacialVelocity, Dornic05:_IntegrationLangevin}.
    This effect is reinforced by the breakdown of the diffusion approximation underlying Eqs.~\eqref{eq:ito-N} and \eqref{eq:ito-x} for small $x$~ \citep{Kessler98:_FrontPropagation}.
    In reality $N$ and so $x$ reach zero eventually, implying global extinction of that species.
    We therefore approximate the dynamics of $x$ by simple Brownian motion with an absorbing boundary at $x=0$  \citep{Pande22:_QuantifyingInvasibility}.

    Now consider the probability that a species starting from $x=x_0$ will still exist after a time $T$, i.e., the probability for $x$ to never reach $0$ before $T$. Textbook methods evaluate this to $\erf(x_0/\sqrt{2 T v_{\mathrm{e}}})$  \citep{gardiner90:_handb_stoch_method}, where $\erf$ denotes the error function  \citep{abramowitz72:_handb_mathem_funct}.
    For $T$ not too near in the future ($T \gg v_{\mathrm{e}}^{-1} x_0^2$), this simplifies to
    \begin{align}
      \label{eq:P-surv}
      (\text{probability of survival until $T$})  = \sqrt{\frac{2}{\pi T v_{\mathrm{e}}}} x_0.
    \end{align}
    That is, for any sufficiently large, fixed observation time $T$, the current value of $x_0=x=\log(1+N/N^*)$ is directly proportional to the probability of species survival.  In Appendix~\ref{sec:numer-test-equat} we demonstrate validity of Eq.~\eqref{eq:P-surv} for a model with discrete population sizes $N$. 

    Data suggest that $v_{\mathrm{e}}$ does not usually vary strongly within taxonomic groups \citep{Kalyuzhny14:_NicheNeutrality, Kessler15:_NeutralDynamics}.
    When this is so, the quantity $\mathcal{L}_{\text{reg}}=\sum_{i}^S\log\left(1+N_i/N_i^*\right)$ is, for a given group of $S$ species, proportional to the expected number of species surviving after a long time $T$.
    
    Above model invokes what is known as an $R$ vortex, one of four mechanisms leading to extinction \citep{Carlson19:_MathematicsExtinction}. Corresponding results for other types of extinction vortices are likely to have a similar structure but use different values for $N_i^*$.
    \end{multicols}
\end{floatbox}

In Box~\ref{box:regularisation} we show for populations driven by environmental and demographic stochasticity that the quantity
\begin{align}
  \label{eq:L-new} \mathcal{L}_{\text{reg}}=\sum_{i}^S\log\left(1+N_i/N_i^*\right)
\end{align}
is, for a given group of $S$ species, approximately proportional to the expected number of species in that group that survive globally after a long time.
(We write $\log(\cdot)$ for natural logarithms.)
Larger values of $\mathcal{L}_{\text{reg}}$ imply that fewer species will go extinct.
Consistent with intuition, an extinct species is treated in Eq.~\eqref{eq:L-new} in the same way as a non-existent species [since $\log(1+N_i/N_i^*)=0$ when $N_i=0$].
Next, we establish a relation between this result and the LPI.

\subsection{The Living Planet Index}
\label{cha:living-planet-index}

The global Living Planet Index (LPI) is designed to track average ``species population trends'' \citep{WWF22:_LivingPlanet} for a given taxonomic or functional group of species.
In defining the LPI, one needs to distinguish the quantity it aims to represent conceptually and how it is being computed in practice.

\subsubsection{Definition of the Living Planet Index}
\label{sec:definition}

Conceptually, LPI represents the geometric mean (defined for $n$ numbers $x_1$, $\ldots$, $x_n$ as $\sqrt[n]{x_1\cdot \ldots \cdot x_n}$) of the global abundances of all species in the group considered, normalised to a fixed baseline year \citep{Collen09:_MonitoringChange}.
Geometric mean abundance biodiversity indices of this type stand out by their combined simplicity, favourable statistical properties, intuitive accessibility and ecological plausibility \citep{Santini17:_AssessingSuitability}.

Mathematically, if $S$ is the total number of species in the group and $N_i$ the global population size of the $i$-th species in this group, one can compute
\begin{align}
  \label{eq:LPI}
  \text{LPI}= \exp
  \left(
  \frac{\mathcal{L}-\mathcal{L}_0}{S}
  \right),
\end{align}
where 
\begin{align}
  \label{eq:L}
  \mathcal{L}=\sum_i^S \log N_i
\end{align}
is the sum of logarithmic population sizes and $\mathcal{L}_0$ the value of $\mathcal{L}$ in the baseline year.

\subsubsection{Computation of the Living Planet Index}
\label{sec:comp-living-plan}

\begin{figure}[t]
  \centering
  \includegraphics[width=0.6\linewidth]{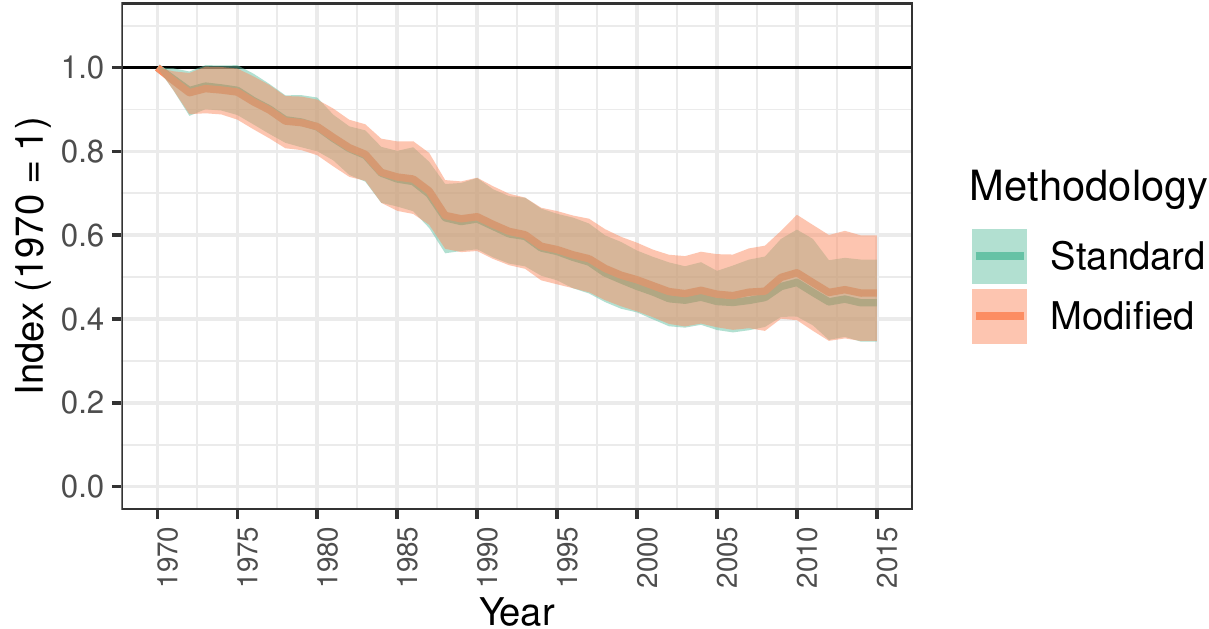}
  \caption{\label{fig:LPI-comparison} [Colour in print] Comparison of the global Living Planet Index for two methods to estimate global abundances trends of species from limited local abundance data. Shaded areas indicate 
    confidence intervals.
    The `Standard' method represents the global abundance trends by the trend of the geometric mean of local abundances, the `Modified' method, described in Sec.~\ref{sec:comp-living-plan}, estimates the trend from the sum of the available local population time series.
    Based on the public subset of the Living Planet Database \citep{LPI22:_LivingPlanet}.}
\end{figure}

The global LPI for vertebrate species is regularly published by the World Wildlife Fund \citep{WWF22:_LivingPlanet}.
The baseline year is 1970.
Its value is estimated from a large database of population time series using a methodology developed by the Zoological Society of London \citep{Collen09:_MonitoringChange}.
The current methodology compensates for incomplete and uneven temporal, taxonomic and geographic coverage by the database \citep{McRae17:_DiversityweightedLiving}. It also takes into account that many time series in the database refer to local or regional rather than global populations.

To address the last issue, the methodology estimates the trend in the global abundance of a recorded species by the trend in the geometric mean of all population time series available for that species.
One might therefore argue that the published LPI does not actually estimate changes in the geometric mean of global population sizes, but the geometric mean of the sizes of local populations, which is different, especially in cases of local species extinctions.
However, we found that the following modification of the methodology, which avoids this issue, changes global LPI estimates only little. Crucially, this variant methodology differs from that of \cite{McRae17:_DiversityweightedLiving}  by replacing the geometric mean in the estimation of a species' global population trend by an arithmetic mean.

After the smoothing and interpolation of all population time series data for a given species following the established methodology \citep{McRae17:_DiversityweightedLiving}, we standardised all these time series to attain the same maximum value. This was done to account for ignorance of the absolute population size that each time series effectively represents.
For each pair of subsequent years $y$, $y+1$, we then first determined the set of those time series for that species that were available in both years (time series are considered available over the time interval over which data have been recorded).
Then we computed for both $y$ and $y+1$ the sums over all these population times series and from the ratio of these two sums the estimated change in $\log_{10}$ global abundance between $y$ and $y+1$.
These estimated increments in $\log_{10}$ global abundance were then averaged over species and regions following the established methodology to compute yearly changes in $\log_{10} \text{LPI}$ \citep{McRae17:_DiversityweightedLiving}.

As we show in Fig.~\ref{fig:LPI-comparison}, the impact of this conceptual refinement of methodology on calculated LPI time series is minimal.
One can therefore safely interpret the published global LPI as estimating trends in global population sizes of species, as given by Equations~\eqref{eq:L} and~\eqref{eq:LPI}.
Our analysis relies on this interpretation.

\subsubsection{Regularisation of the Living Planet Index and its link to extinction risk}
\label{sec:extinction-risk}

If one of the $S$ species entering the LPI goes extinct ($N_i=0$) at some time after the baseline year, the quantify $\mathcal{L}$ defined by Eq.~\eqref{eq:L} attains a value of negative infinity for that year and LPI becomes zero by Eq.~\eqref{eq:LPI}, irrespective of all other species.
Practical calculations of the LPI avoid this mathematical singularity by introducing \textit{ad hoc} rules to handle rare or extinct species or populations \citep{Collen09:_MonitoringChange}.
So far, satisfactory resolution of this issue has been hampered by insufficient understanding of geometric mean abundance metrics in terms of ecological first principles.

Here we propose to resolve this issue by making use of the intuitive interpretation of $\mathcal{L}_{\text{reg}}$, defined by Eq.~\eqref{eq:L-new}, as quantifying long-term species survival.
Using $\mathcal{L}_{\text{reg}}$ in place of $\mathcal{L}$ and correspondingly defining the \emph{regularised Living Planet Index} as
\begin{align}
  \label{eq:LPI-new}
  \text{LPI}_{\text{reg}}= \exp
  \left(
  \frac{\mathcal{L}_{\text{reg}}-\mathcal{L}_{\text{reg},0}}{S}
  \right),
\end{align}
with $\mathcal{L}_{\text{reg},0}$ denoting the value of $\mathcal{L}_{\text{reg}}$ at the baseline year, the singularities occurring when species go extinct disappear. We note that, to avoid a shifting baseline \citep{Pauly95:_AnecdotesShifting}, neither $S$ nor $\mathcal{L}_{\text{reg},0}$ should exclude species that went extinct since the baseline year.

Otherwise, when all species populations $N_i$ are much larger than the corresponding $N_i^*$, $\text{LPI}$ and $\text{LPI}_{\text{reg}}$ are nearly identical.
Both are then strictly increasing functions of mean long-term species survival probability in our approximation.
This result should settle the debate between \cite{Pereira06:_GlobalMonitoring} and \cite{Collen09:_MonitoringChange} about the extent to which the LPI is a metric of extinction risk.


\subsection{Metacommunity modelling}

To study the implications of metacommunity structure and dynamics for the use of BICs, we employed the Lotka-Volterra Metacommunity Model (LVMCM, \citealt{OSullivan19:_Metacommunity-scaleBiodiversity,OSullivan21:_IntrinsicEcological, OSullivan23:_CommunityComposition}).
The model simulates the dynamics of population biomasses in species-rich metacommunities, modelled as spatial networks of Lotka-Volterra competition models, each understood to represent a distinct `patch' in space, coupled by dispersal.
Competition between species leads to biotic filtering, which controls community composition in addition to abiotic filtering modelled by varying intrinsic growth rates between species and patches.
Metacommunities are assembled to saturation, i.e., by iteratively introducing new random species at low abundance until each such addition generates one extinction on average.
Previous work has shown that, as they saturate, these metacommunities self-organise to reproduce a wide range of well-documented macroecological patterns \citep{OSullivan19:_Metacommunity-scaleBiodiversity,OSullivan21:_IntrinsicEcological}.

For the detailed model structure, see \cite{OSullivan23:_CommunityComposition}.
In the present study we simulated metacommunities with 100 patches randomly distributed over a $10\times 10$ square, with dispersal length $0.5$, and set the niche width parameter of \cite{OSullivan23:_CommunityComposition} to $0.2$.
Intraspecific competition strength was $1$ for all species, and any species suppressed any given other species with probability $0.3$, in which case the corresponding interaction strength was sampled from a beta distribution with shape parameters $\alpha=\beta=2$.
The resulting metacommunities contained about 320 species and sustained at each patch around 50 species with population size larger than a hundredth of the single-species carrying capacity.

\section{Definition and theory of Biodiversity Impact Credits}
\label{sec:theory}

We now propose our biodiversity impact metric BIC and report some of its key characteristics, especially in the context of a biodiversity credit market.

\subsection{Quantifying local biodiversity impact}
\label{sec:local-biod-impact}

Since $\mathcal{L}_{\text{reg}}$ is proportional to the expected number of surviving species in our approximation, we can quantify the impact of any human intervention on long-term species extinction risk by the resulting change $\Delta \mathcal{L}_{\text{reg}}$ in this metric.
If the changes in population sizes $\Delta N_i$ resulting from this impact are small compared to total population sizes $N_i$, as will often be the case, $\Delta \mathcal{L}_{\text{reg}}$ is well approximated by a Taylor expansion to linear order in $\Delta N_i$, i.e., as
\begin{align}
  \label{eq:delta-L-reg}
  \Delta \mathcal{L}_{\text{reg}} \approx \sum_i^S \frac{\partial \mathcal{L}_{\text{reg}}}{\partial N_i} \Delta N_i= \sum_i^S \frac{\Delta N_i}{N_i^*+N_i}.
\end{align}
The notation $\partial \mathcal{L}_{\text{reg}}/\partial N_i$ denotes the first derivative of $\mathcal{L}_{\text{reg}}$ with respect to $N_i$, computed while keeping all other $N_j$ (with $j\ne i$) fixed. 
From Eq.~\eqref{eq:delta-L-reg} it is clear that $\Delta \mathcal{L}_{\text{reg}}$ weights impacts on globally rare species higher than impacts on common species, thus plausibly providing an appropriate measure of pressure on ($\Delta \mathcal{L}_{\text{reg}}<0$), or relief to ($\Delta \mathcal{L}_{\text{reg}}>0$) global biodiversity.

\subsection{The Biodiversity Impact Credit metric}
\label{sec:biod-gain-cred}

In view of above result, we define the \emph{Biodiversity Impact Credits} associated with an area or site $\alpha$ as
\begin{align}
  \label{eq:BIC}
  \text{BIC}_\alpha=\sum_i^S \frac{\Delta n_{\alpha,i}}{N_i^*+N_i},
\end{align}
with $N_i$ denoting the \emph{current} global abundance of species $i$ and $\Delta n_{\alpha,i}$ denoting the difference between the \emph{current} local abundances and a local baseline abundance documented at some time in the past.
While sometimes we refer to $\alpha$ as `land' held by an organisation, it could equally be an area of aquatic habitat.
The $\Delta n_{\alpha,i}$ do not need to be small compared to $N_i$ and can be negative.

For values of $N_i$ larger than about 100-1000 individuals, $N_i^*$ is generally sufficiently small to be disregarded in the evaluation of Eq.~\eqref{eq:BIC} \citep{Melbourne12:_StochasticityDemographic, Kalyuzhny14:_NicheNeutrality}.
The $N_i^*$ thus play the role of regularisation constants relevant only for species close to global extinction.
Where required, $N^*$ values can be computed as $N^*=v_{\mathrm{d}}/v_{\mathrm{e}}$, with $v_{\mathrm{d}}$ and $v_{\mathrm{e}}$ estimated from population time series data as used in calculations of the global LPI \citep{Kalyuzhny14:_NicheNeutrality, Kessler15:_NeutralDynamics}, life-history data \citep{Saether04:_LifeHistory,Saether13:_HowLife, engen09:_reprod_value_and_stoch_demog}, or combinations thereof \citep{Lande03:_DemographicEnvironmental, Saether09:_CriticalParameters}.
Note, however, that the value of $v_{\mathrm{e}}$ may depend on extent and location of species ranges \citep{Lande99:_SpatialScale}.
Considering that species close to extinction tend to have comparatively narrow ranges \citep{Mace08:_QuantificationExtinction}, we recommend in such cases to determine a value of $v_{\mathrm{e}}$ that correspond to the location and range of the near-extinct species in question.

For the use of BICs in a biodiversity credit markets, we propose a set of rules that is designed to counteract the market incentive to maximise BICs by artificially minimising baseline abundances:
\begin{enumerate}
\item  Tradable gains can be claimed only against empirically determined published baseline abundances.\label{item:1}
\item When publishing baseline abundances to determine BICs for a site, values for all species within the group considered must be published at the same time.
  For species that are not observed in the baseline year the baseline value is the first empirically non-zero abundance.\label{item:2}
\item Baseline abundances can be amended when more accurate data become available but not to get a `fresh start' after a population decline.\label{item:3}
\end{enumerate}
Rule~\ref{item:1} prevents baselines and so gains in BICs from being manipulated based on speculative assumptions.
Rule~\ref{item:2} forbids the choice of different baseline years for different species in order to artificially inflate BICs.
The exception for unobserved species that we included takes account of the possibility that empirical methods and so detectability improves over time, so that low-abundance species that were initially missed are later observed despite having stable population.
It is the conservative choice (generates smaller BICs) compared to the alternative of setting baseline abundance of unobserved species to zero.
Without Rule~\ref{item:3} land-holders would be tempted to erase the memory of past population declines by re-setting baselines when populations are minimal.
We recognise, though, that these rules are not perfect and that there may be real-world situations where either stricter or laxer rules for baseline setting are reasonably justified.

Two desirable properties of BICs follow directly from Eq.~\eqref{eq:BIC}: First, BICs are additive, that is, the BICs of a combination of non-overlapping areas is simply the sum of their BICs (Appendix~\ref{sec:non-additv-chang}).
Additivity is a crucial property for a tradable credit metric. If violated, credits can appear or disappear by simply merging or splitting the areas of sites in paper transactions without actual ecological change---an undesirable outcome.
We note that the quantification of biodiversity gain (or loss) in terms of the exact effects local populations gains (or losses) $\Delta n_{\alpha,i}$ have on $\mathcal{L}_{\text{reg}}$, i.e.\
\begin{align}
  \label{eq:DeltaL}
  \Delta\mathcal{L}_{\text{reg}} =
  \sum_{i}^S\log\left[1+N_i/N_i^*\right]-\log\left[1+(N_i-\Delta
  n_{\alpha,i})/N_i^*\right]=  \sum_{i}^S\log\left[\frac{N_i^*+N_i}{N_i^*+N_i-\Delta
  n_{\alpha,i}}\right],
\end{align}
is \emph{not} additive in this sense (Appendix~\ref{sec:non-additv-chang}).
This makes the approximate measure of impact on extinction risk BIC preferable over the exact measure $\Delta\mathcal{L}_{\text{reg}}$ in a market context.

Second, BICs represent a strictly finite resources. The total population sizes $N_i$ of each species $i$ can never be smaller than the sum of population increments over all sites $\sum_\alpha \Delta n_{\alpha,i}$.
The sum of all BIC
\begin{align}
  \sum_\alpha \text{BIC}_\alpha=\sum_\alpha \sum_i^S \frac{\Delta n_{\alpha,i}}{N_i^*+N_i}=\sum_i^S \frac{\sum_\alpha \Delta n_{\alpha,i}}{N_i^*+N_i}
\end{align}
is therefore always less than $S$.
It approaches $S$ in the hypothetical case of large increments $\sum_\alpha \Delta n_{\alpha,i}$.
Numerical BIC values can thus be interpreted by comparison to global species richness $S$, implying that BICs have units of `species'.

\subsection{Incentive structure generated by BICs}
\label{sec:incentive-structure}

We now analyse the market forces generated by BICs in a perfect market, asking in particular whether the metric might generate perverse market incentives.
Let $p$ be the market price of one BIC and $C_{\alpha}=C_{\alpha}(n_{\alpha,1},\ldots,n_{\alpha,S})$ the net present value \citep{Vernimmen14:_CorporateFinance} of the costs of sustaining local population sizes $n_{\alpha,1},\ldots,n_{\alpha,S}\ge 0$ in areas of land held by market participant $\alpha$, including costs related to lost opportunities for other uses, minus the net present value of the resulting ecosystem services.
Denote by $n_{\alpha,i,0}$ the baseline abundances for species $i$ over all sites held by market participant $\alpha$, so that $\Delta n_{\alpha,i}=n_{\alpha,i}-n_{\alpha,i,0}$ are the increments entering Eq.~\eqref{eq:BIC}.
The total value of that area is then $V_{\alpha} = -C_{\alpha}(n_{\alpha,1},\ldots,n_{\alpha,S})+ p\,\text{BIC}_{\alpha}$, plus value unrelated to $n_{\alpha,1},\ldots,n_{\alpha,S}$.

In a perfect market, market participant $\alpha$ will take measures to maximise $V_{\alpha}$, at which point
\begin{align}
  \label{eq:local-rule}
  0=\frac{\partial V_{\alpha}}{\partial n_{\alpha,i}}= -\frac{\partial C_{\alpha}}{\partial n_{\alpha,i}} + \frac{p}{N^*_i+N_i}- \frac{p \,\Delta n_{\alpha,i}}{\left(N^*_i+N_i\right)^2}\quad
  (\text{or}\quad  n_{\alpha,i}=0)
\end{align}
for all species $1\le i \le S$. The last term in Eq.~\eqref{eq:local-rule} results because $n_{\alpha,i}$ contributes to total population size $N_i$(so $\partial N_i/\partial n_{\alpha,i}=1$).
The $n_{\alpha,i}=0$ case arises when there are no net benefits at all in sustaining species $i$.
In such a case one typically finds that even small populations of $n_{\alpha,i}>0$ do not add value, implying that $\partial C_{\alpha}/\partial n_{\alpha,i} > p/(N^*_i+N_i) + p \,n_{\alpha,i,0}/\left(N^*_i+N_i\right)^2$ at $n_{\alpha,i}=0$, which in turn implies that $\partial C_{\alpha}/\partial n_{\alpha,i} > p/(N^*_i+N_i)$.



Now, consider the problem of minimising the global costs of sustaining $\mathcal{L}_{\text{reg}}$ at a given level by appropriately choosing the population sizes $n_{\alpha,i}\ge 0$ sustained by each land holder $\alpha$.
Solutions of such a constrained non-linear optimisation problem satisfy the Karush–Kuhn–Tucker conditions \citep{Chiang05:_FundamentalMethods}.
These are derived from a Lagrangian, which here becomes
\begin{align}
  \label{eq:goal-func}
  \sum_\alpha C_{\alpha} - \sum_{\alpha,i}\mu_{\alpha,i}n_{\alpha,i} -\lambda \mathcal{L}_{\text{reg}},
\end{align}
where $\mu_{\alpha,i}$ and $\lambda$ are the Karush-Kuhn-Tucker multipliers.
The resulting Karush–Kuhn–Tucker conditions are that
\begin{align}
  \label{eq:stationarity}
  -\frac{\partial C_{\alpha}}{\partial n_{\alpha,i}} - \mu_{\alpha,i}+ \frac{\lambda}{N^*_i+N_i} = 0
\end{align}
for all $\alpha$ and $i$, where either $\mu_{\alpha,i}=0$ and $n_{\alpha,i}\ge 0$ or $\mu_{\alpha,i}>0$ and $n_{\alpha,i}=0$.
Comparison of Eqs.~\eqref{eq:local-rule} and \eqref{eq:stationarity} and of the considerations for $n_{\alpha,i}=0$ shows that with $p=\lambda$ the two conditions are identical, except for the term ${p \Delta n_{\alpha,i}}/{\left(N^*_i+N_i\right)^2}$ in Eq.~\eqref{eq:local-rule}.
However, this term makes a sizeable contribution compared to the term $p/(N^*_i+N_i)$ only when $\alpha$ has changed the abundance of a species $i$ by an amount that is comparable to or larger than the current global abundance $N_i$ (so $\Delta n_{\alpha,i}$ and $N_i$ are of comparable size).
Absent such dominant market participants, a perfect BIC market (arising with many omniscient, perfectly rational, profit-maximising actors) leads to near optimal allocation of resources to sustain $\mathcal{L}_{\text{reg}}$, and so $\text{LPI}_{\text{reg}}$, at a given level.
Larger $\text{LPI}_{\text{reg}}$ correspond to higher BIC prices $p$.

To study the case of dominant market participants, we first consider the case of negative $\Delta n_{\alpha,i}$.
In this case, Eq.~\eqref{eq:local-rule} can be read as implying that the marginal cost ${\partial C_{\alpha}}/{\partial n_{\alpha,i}}$ that $\alpha$ is willing to incur to maintain the abundance of species $i$ is by a factor $1+|\Delta n_{\alpha,i}|/\left(N^*_i+N_i\right)$ larger than the price of the resulting BICs\@.
This additional conservation effort will not usually be detrimental.

The opposite case, positive $\Delta n_{\alpha,i}$ that are of similar magnitude as $N_i$, can arise only when a large fraction of the population of species $i$ is held by $\alpha$.
To address such situations we now show that, firstly, even when a species is dominantly held by a single market participant, BICs still incentivise protecting this species and growing its populations and, secondly, BICs disincentivise dominance.

\begin{figure}[t]
  \center{}
  \includegraphics[width=0.7\linewidth]{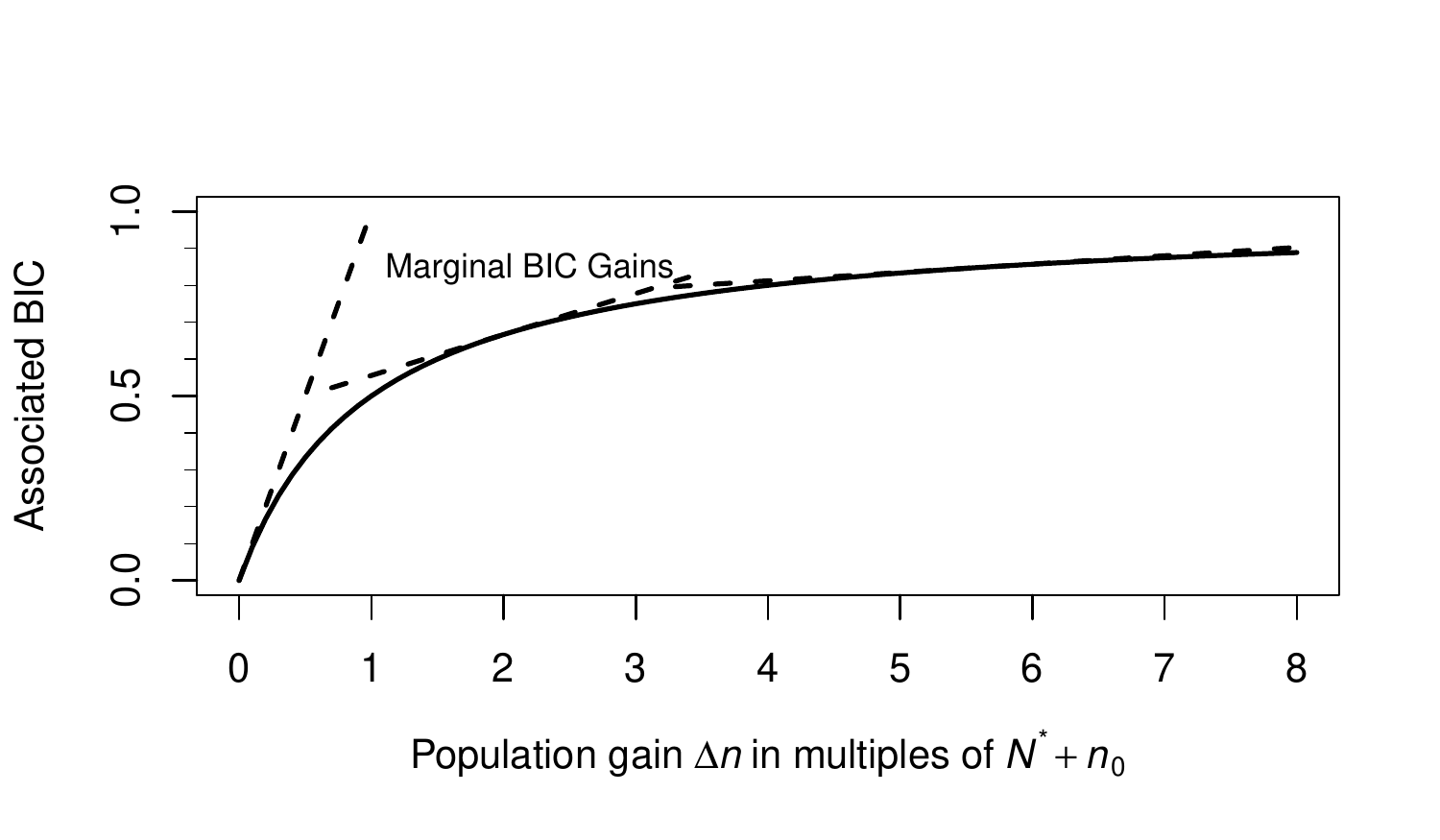}
  \caption{\textbf{Accumulation of BICs when rebuilding a species as a dominant market participant.}
     [no colour in print] The solid line gives the BICs associated with this species, Eq.~\eqref{eq:monopoly-bsc}.
    As illustrated by the dashed lines, BICs gained by increasing the species' population by a single individual are largest when it is closest to its baseline population $n_0$.
    However, to achieve full BIC, the species' population must be lifted well above $N^*+n_0$.\label{fig:extinction-BIC}}
\end{figure}

Consider first a market participant $\alpha$ whose land is home to the entire global population of some species $i$.
Suppressing the indices $i$ and $\alpha$,
assume that this market participant has grown this population from an initial size of $n_{0}$, so that $\Delta n=N-n_{0}$, implying $N=n_{0}+\Delta n$.
The participant's BICs associated with species $i$ are then
\begin{align}
  \label{eq:monopoly-bsc}
  \text{associated BICs}=\frac{\Delta n}{N^*+n_{0}+\Delta n}.
\end{align}
A population gain of $\Delta n=N^*+n_{0}$, for example, earns this market participant BICs worth $0.5$ species (Fig.~\ref{fig:extinction-BIC}). For a species that was originally close to extinction, so that $n_{0}$ is of similar magnitude as $N^*$, maintaining the population at this or even higher levels can be well worth while, even when, by Eq.~\eqref{eq:local-rule}, the effective value of BICs for this species is for $\alpha$ discounted by a factor $1-\Delta n/(N^*+n_{0}+\Delta n)=(N^*+n_{0})/(N^*+n_{0}+\Delta n)$.
Importantly, market incentives to rebuild the species' population are highest for the initial gains (Fig.~\ref{fig:extinction-BIC}).

To see that BICs disincentivise market dominance, note that market participants that achieved a smaller population gain of a species $i$ on their land benefit more from increasing this population further than those who have achieved larger gains.
The reason is that increasing a population not only generates further BIC gains through the numerator in Eq.~\eqref{eq:BIC} but also penalises the value of previous gains by increasing global species abundance in the denominator, and this penalty is larger for those who have achieved larger previous gains.
This disincentivises dominant holdings of population gains.
We therefore expect that market misalignment due to the difference between conditions Eq.~\eqref{eq:local-rule} and~\eqref{eq:stationarity} will be harmless in most cases, implying near-optimal resource allocation as explained above.

\subsection{Consideration of off-site impacts}
\label{sec:spec-divers-impact}

For most business activities, impacts on biodiversity are not constrained to sites the businesses hold.
Often, these off-site impacts will be spread out widely, e.g., along complex supply chains or because they result from widely dispersing pollutants.
(Off-sites effect due to dispersal of populations are considered in Sec.~\ref{sec:local-vs-regional}.)
It can be impractical for a business to fully avoid the resulting non-local impacts on species populations.

In order not to endanger, despite this, attainment of the societal goal of reducing species extinction risk, a business (or a similar organisations) may choose to compensate the non-local increase in mean long-term species extinction risk resulting from its activity by generating BICs at sites it holds or by purchasing BICs that other organisations generate for this purpose.

Specifically, denote by $\Delta N_{\text{off},i}$ the diffuse off-site changes in species abundances generated by the activity of a business compared to the baseline, and define
\begin{align}
  \label{eq:BIC-off}
  \text{BIC}_{\text{off}}=\sum_i^S\frac{\Delta N_{\text{off},i}}{N_i^*+N_i},
\end{align}
with $N_i$ again denoting current species abundances.
Since the $\Delta N_{\text{off},i}$ will usually be negative, so will $\text{BIC}_{\text{off}}$.
However, we can show that when
\begin{align}
  \label{eq:offsetting-equation}
\text{BIC}_\alpha + \text{BIC}_{\text{off}} > 0
\end{align}
the resulting change in $\mathcal{L}_{\text{reg}}$ is always positive.
This result holds for both small and large $\text{BIC}_\alpha$ and $\text{BIC}_{\text{off}}$ without invoking any approximation.
Since a positive change $\Delta\mathcal{L}_{\text{reg}}$ implies a positive overall impact on mean long-term species survival, one can identify Eq.~\eqref{eq:offsetting-equation} as a condition for a business to be biodiversity positive with respect to protection of species.

The proof of above result starts with the analogue of Eq.~\eqref{eq:DeltaL}, taking both on-site and off-site changes in abundances into account, and then makes use of the fact that $\log(1+x) \le x$ for any $x>-1$ (so that $-\log(1+x) \ge -x$) and of Eq.~\eqref{eq:offsetting-equation} to demonstrate an increase in $\mathcal{L}_{\text{reg}}$:
\begin{align}
  \label{eq:offsetting-theorem}
  \begin{split}
    \Delta \mathcal{L}_{\text{reg}}&=
    \sum_{i}^S\log\left[1+N_i/N_i^*\right]-\log\left[1+(N_i-\Delta
      n_{\alpha,i}-\Delta N_{\text{off},i})/N_i^*\right]\\
    &=
    \sum_{i}^S\log\left[1+N_i/N_i^*\right]-\log\left[\left(1+N_i/N_i^*\right)\left(1-
        \frac{(\Delta n_{\alpha,i}+\Delta N_{\text{off},i})/N_i^*}{1+N_i/N_i^*}\right)\right]\\
    &=
    \sum_{i}^S-\log\left[1-
        \frac{(\Delta n_{\alpha,i}+\Delta N_{\text{off},i})/N_i^*}{1+N_i/N_i^*}\right]\\
    &=-\sum_{i}^S\log\left[1-
       \frac{\Delta n_{\alpha,i}+\Delta N_{\text{off},i}}{N_i^*+N_i}\right]\\
    &\ge \sum_{i}^S
     \frac{\Delta n_{\alpha,i}+\Delta N_{\text{off},i}}{N_i^*+N_i} \\
  &= \sum_{i}^S
  \frac{\Delta n_{\alpha,i}}{N_i^*+N_i} + \sum_{i}^S
  \frac{\Delta N_{\text{off},i}}{N_i^*+N_i} \\
  &= \text{BIC}_{\alpha} + \text{BIC}_{\text{off}} >0
\end{split}
\end{align}

\subsection{The Biodiversity Stewardship Credit metric}
\label{sec:bsc}

A special case of the BIC metric arises when the baseline abundances of all species are zero.
Such a situation might occur, e.g., when rewilding barren land.
From the abstract standpoint that all land and water on Earth was originally lifeless, one can argue that zero is the natural value for all baseline abundances, and that the current holders of any site are entitled to credits for sustaining, to the present day, the biodiversity on this site passed on to them by their predecessors.
The credits are then awarded not for changes in biodiversity but for exercising stewardship over existing biodiversity.
We therefore call the variant of BICs where the baseline abundances are set to zero Biodiversity Stewardship Credits (BSC).
The explicit formula for this metric is
\begin{align}
  \label{eq:BSC}
  \text{BSC}_\alpha=\sum_i^S \frac{n_{\alpha,i}}{N_i^*+N_i},
\end{align}
with $n_{\alpha,i}$ denoting the sustained abundance of species $i$ at site $\alpha$.

BSCs share with BICs all the properties derived above.
While BSCs have the advantage over BICs of being conceptually simpler, BSCs are not immediately suitable for a trade in biodiversity impacts.
Global BSCs are overabundant, because, disregarding the regularisation constants $N^*_i$, the global sum of all BSCs equals global species richness $S$.
Trading BSCs off against negative impacts could thus theoretically bring Earth back to its lifeless primordial state.

BSCs have their own role to play when biodiversity stewardship itself is of interest rather than the balancing of impacts.
Organisations can, for example, include accounts of the total BSCs they hold in yearly reports to demonstrate their biodiversity credentials.

A second important role for BSC arises because BICs can often be approximated by the change in BSCs between baseline year and present:
\begin{align}
  \label{eq:BIC-vs-BSC}
  \text{BIC} \approx \Delta \text{BSC}.
\end{align}
To quantify the accuracy of this approximation, we compute the error in Eq.~\eqref{eq:BIC-vs-BSC} for the contribution from a single species $i$ in comparison with this species' contribution to current $\text{BSC}_{\alpha}$.\@
Denoting by $N_{i,0}$ the species' global abundance in the baseline year, we obtain
\begin{align}
  \label{eq:error}
  \left|
  \frac{
  \left(
  \frac{n_{\alpha,i}}{N_i^*+N_i} - \frac{n_{\alpha,i,0}}{N_i^*+N_{i,0}}
  \right)-\frac{\Delta n_{\alpha,i}}{N_i^*+N_i}}{\frac{n_{\alpha,i}}{N_i^*+N_i}}
  \right|=\frac{n_{\alpha,i,0}
  \left|
  N_i-N_{i,0}
  \right|}{n_{\alpha,i}
  \left(
  N_i^*+ N_{i,0}
  \right)}.
\end{align}
This relative error is small compared to one, e.g., when the species' local baseline abundance $n_{\alpha,i,0}$ was not much larger than its current abundance $n_{\alpha,i}$ and the absolute change in the species' global population size $\left| N_i-N_{i,0} \right|$ is small compared to the global baseline abundance $N_{i,0}$. While these conditions will often be satisfied, they break down, e.g., when a site holder rebuilds a species' population from the verge of extinction.
In such cases care should be taken to use BICs rather than BSCs to avoid undesirable artefacts.
Thus, while site holders can often track changes in their BSCs to estimate their BICs, they should keep records of baseline abundances $n_{\alpha,i,0}$ so BICs can accurately be computed using Eq.~\eqref{eq:BIC} if required.




Since BICs and BSCs are additive across sites one can define a corresponding \emph{BIC density} and \emph{BSC density} by dividing BICs and BSCs, respectively, by site area, thus making metric values more comparable across sites of different size.
Below we will provide some empirical values of this density in units of $\mathrm{species}\,\mathrm{km}^{-2}$.



\subsection{Local vs regional impacts on species populations}
\label{sec:local-vs-regional}

\begin{figure}
  \centering
  \includegraphics[width=0.5\linewidth]{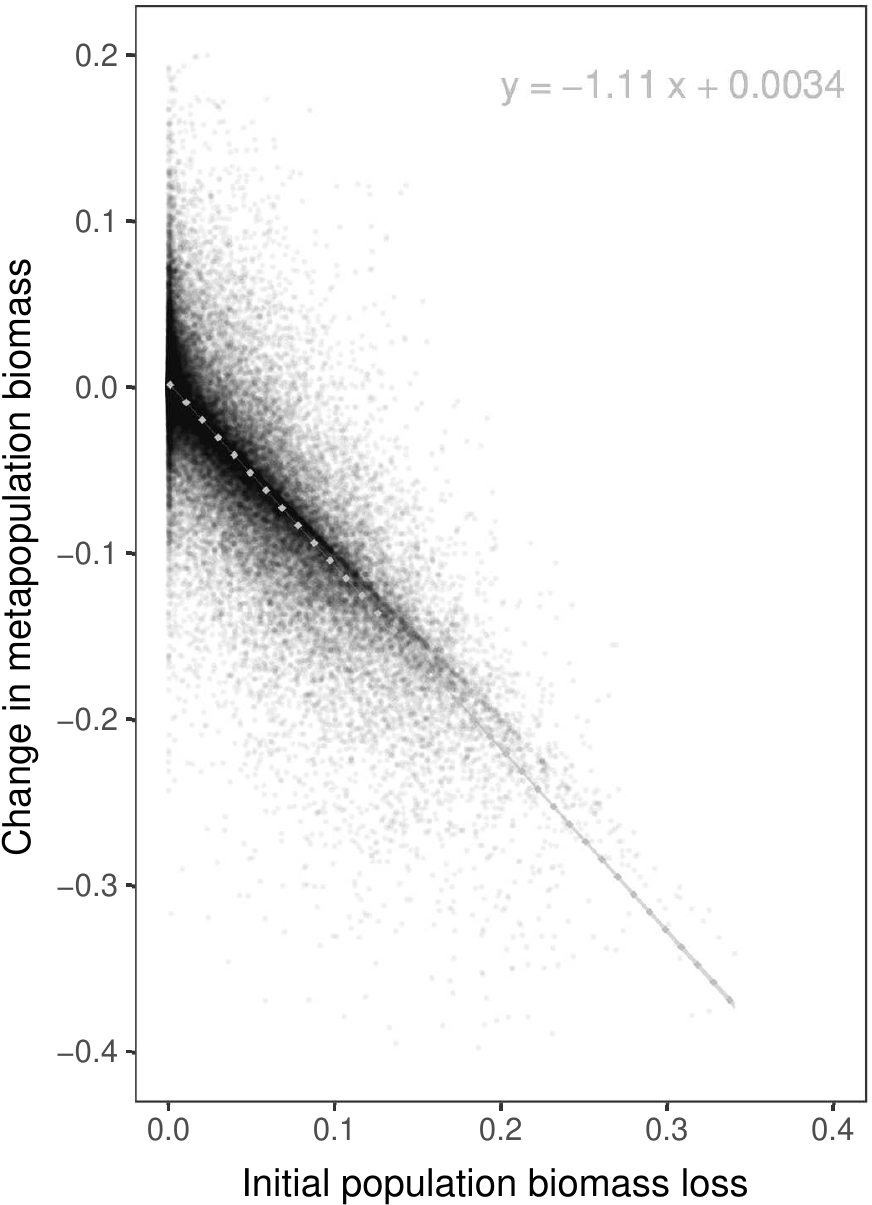}
  \caption{\label{fig:impact}  [no colour in print] \textbf{Long-term, regional-scale impacts of localised perturbations in simulated metacommunities.} In each simulation, we removed a single patch from a model metacommunity and compared the initial population biomasses of each species in that patch with the resulting long-term impact on the regional biomass of that species. While there is variation in the long-term biomass change, in some cases larger than the removed biomass, on average biomass removed is a good predictor of long-term regional biomass loss. Axes are equally scaled.
    The grey dotted line represents a linear regression.
  }
\end{figure}

Both BICs and BSCs are defined in terms of the local population sizes of species at sites, disregarding the effects local changes in these populations might have at regional level.
Such regional responses might take time to unfold \citep{tilman94:_habit_destr_extin_debt,jackson10:_balan,Essl15:_DelayedBiodiversity} and can be difficult to predict.
This might raise concerns that the long-term ecological reverberations of local impacts are yet to unfold after they registered in BICs or BSCs, and so be missed by these metrics.

Indeed, we demonstrated such complex, far-reaching long-term effects of perturbations in recent simulations of the LVMCM \citep{OSullivan23:_CommunityComposition}.
Remarkably, however, these simulations also reveal that, at least in the case of complete eradication of all species on a site (as when building a warehouse on a meadow), the \emph{average} long-term change in the total population of each species across the metacommunity is nearly identical to the size of the local population removed by the intervention (Fig.~\ref{fig:impact}).
This suggests that quantification of biodiversity impacts in terms of local, short-term changes in population sizes provides a good estimate of the expectation value of the long-term impact, even when ecological complexity makes the actual long-term impacts less predictable.

\subsection{Rewilding}
\label{sec:rewilding}

\begin{figure}
  \centering
  \includegraphics{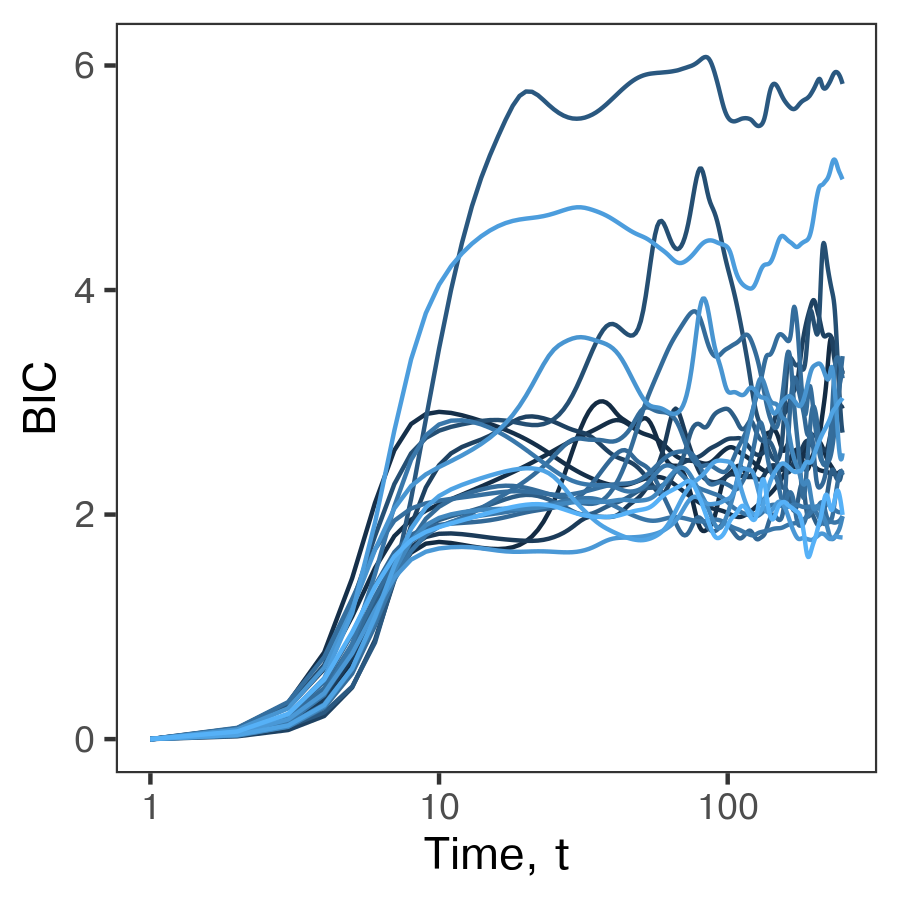}
  \caption{\label{fig:rewild}  [no colour in print] \textbf{Accruement of BICs during local site rewilding.} Each line corresponds to a different LVMCM simulation of `rewilding' a single site as detailed in the text.
    After verifying that the precise value of $N^*$ has little effect on metric values, we set $N^*$ to one $10^{\text{th}}$ of the local single-species carrying capacity.
    Note the logarithmic time axis.
    Accruement tends to be fast but can be followed by ongoing slow fluctuations.
  }
\end{figure}

The perhaps simplest approach to gaining BICs is to permit the natural rewilding of a barren area.
We used LVMCM simulations to get a first idea of how BICs in such areas increase over time as rewilding progresses.
For this, we first kept all biomass at zero in a given site in our metacommunity model and allowed the surrounding metacommunity to relax to a new equilibrium or steady state.
Then we allowed recolonisation of the barren site to occur, recording the associated BICs, where we used as the `global' abundances of species their metacommunity-level population biomasses.
As shown in Fig.~\ref{fig:rewild}, BICs rapidly accumulate over the first $10$ unit times.
Since the intrinsic growth rates of local populations in absence of competitors are $\mathcal{O}(1)$ in our model, these results suggest that BICs tend to recover on a similar time scale as community biomass.
The fluctuations in BICs on longer time scales seen in Fig.~\ref{fig:rewild} are due to intrinsically driven community turnover \citep{OSullivan21:_IntrinsicEcological} and can similarly be expected on real sites where community turnover occurs.

\section{Calculating BICs and BSCs}
\label{sec:calculation}

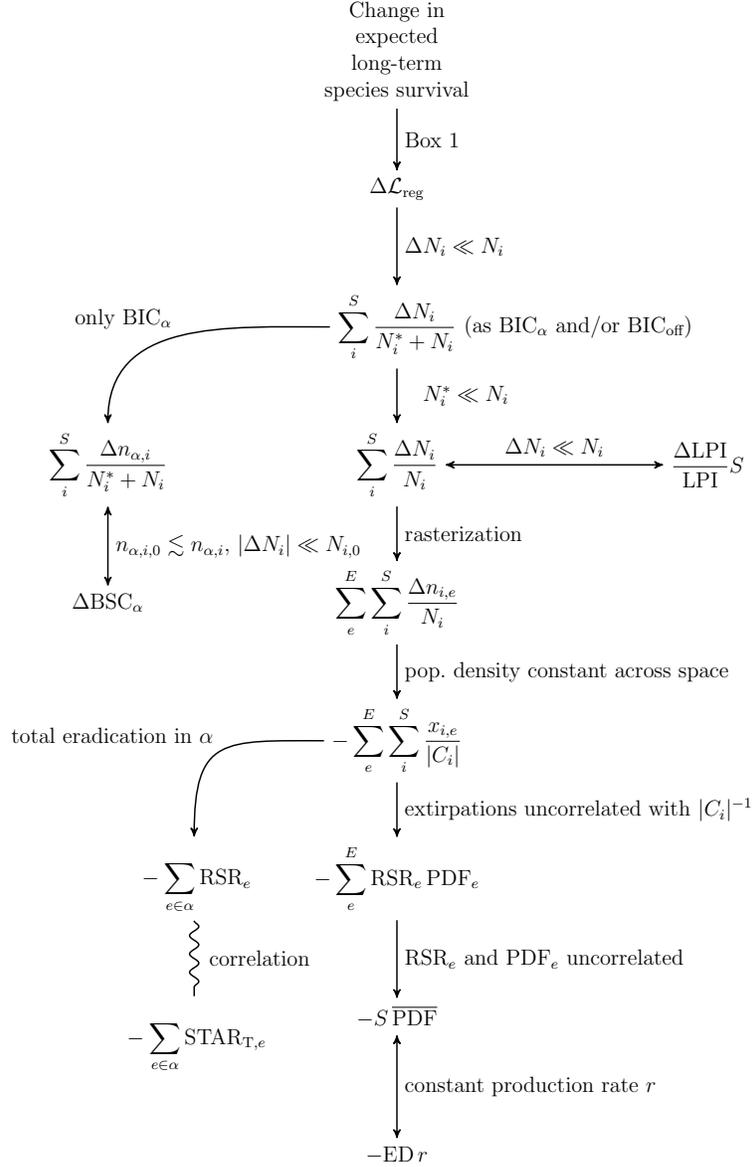
\begin{figure}
  \centering
  \resizebox{!}{.8\textheight}{%
    \begin{tikzpicture}[->,>=stealth',auto,node distance=6.4em,
    thick]
    
    \node (delPsurv) [=1, text centered, text width=3cm] {
      Change in\\ expected long-term species survival
    };
    \node (delLreg) [below of = delPsurv] {$\Delta \mathcal{L}_{\text{reg}}$};
    \node (BIC) [below of = delLreg] {$\displaystyle\sum_i^S\frac{\Delta N_i}{N_i^*+N_i}$};
    \node (BICcomment) [right of = BIC, align=left, xshift=2em] {(as $\text{BIC}_{\alpha}$ and/or $\text{BIC}_{\text{off}}$)};
    \node (delN) [below of = BIC] {$\displaystyle\sum_i^S\frac{\Delta N_i}{N_i}$};
    \node (delLPI) [right of = delN, xshift=8em] {$\displaystyle \frac{\Delta \text{LPI}}{\text{LPI}}S$};
    \node (deln) [below of = delN] {$\displaystyle\sum_e^E\sum_i^S\frac{\Delta n_{i,e}}{N_i}$};
    \node (BICalpha) [left of = delN, xshift=-7em] {$\displaystyle\sum_i^S\frac{\Delta n_{\alpha,i}}{N_i^*+N_i}$};
    \node (delBSC) [below of = BICalpha] {$\displaystyle\Delta \text{BSC}_\alpha$};
    \node (x) [below of = deln] {$\displaystyle -\sum_e^E\sum_i^S\frac{x_{i,e}}{|C_i|}$};
    \node (RSRPDF) [below of = x] {$\displaystyle -\sum_{e}^E \text{RSR}_e\, \text{PDF}_e$};
    \node (RSR) [left of = RSRPDF, xshift=-3em] {$\displaystyle \vphantom{\sum_{e}^E}-\sum_{e\in \alpha} \text{RSR}_e$};
    \node (STAR) [below of = RSR, yshift=-1em] {$\displaystyle \vphantom{\sum_{e}^E}-\sum_{e\in \alpha} \text{STAR}_{\text{T},e}$};
    \node (SDSPDF) [below of = RSRPDF] {$\displaystyle -S\,\overline{\text{PDF}}$};
    \node (EDr) [below of = SDSPDF] {$\displaystyle -\text{ED}\,r$};
    \draw [->] (delPsurv) to node[anchor=west] {Box~\ref{box:regularisation}}  (delLreg);
    \draw [->] (delLreg) to node[anchor=west] {$\Delta N_i \ll  N_i$}(BIC);
    \draw [->] (BIC) to node[anchor=west] {$\phantom{\Delta}N_i^* \ll N_i $} (delN);
    \draw [->] (delN) to node[anchor=west] {rasterization} (deln);
    \draw [->] (deln) to node[anchor=west] {pop.\ density constant across space} (x);
    \draw [->] (x) to node[anchor=west] {extirpations uncorrelated with $|C_i|^{-1}$} (RSRPDF);
    \draw [->] (RSRPDF) to  node[anchor=west] {$\text{RSR}_e$ and $\text{PDF}_e$ uncorrelated} (SDSPDF);
    \draw [<->] (SDSPDF) to  node[anchor=west] {constant production rate $r$} (EDr);
    \draw [<->] (BICalpha) to node[anchor=west] {$n_{\alpha,i,0}\lesssim n_{\alpha,i}$, $|\Delta N_i|\ll N_{i,0}$} (delBSC);
    \draw [<->] (delN) to node[anchor=south]{$\Delta N_i \ll N_i$} (delLPI);
    \draw [->] (x) to[out=180,in=90,looseness=1.6] node[anchor=south east] {total eradication in $\alpha$} (RSR);
    \draw [-,decorate, decoration={snake}] (RSR) to  node[anchor=west] {{ }correlation} (STAR);    
    \draw [->] (BIC) to[out=180,in=90,looseness=1.1]  node[anchor=south east] {only $\text{BIC}_\alpha$} (BICalpha);
  \end{tikzpicture}%
  }
  \caption{Alternative approximate determinations of BICs. All formulas provide approximately the same value in units of `species'.
    Notation and approximations are discussed in Section \ref{sec:calculation}. Annotations of arrows indicate the simplifying assumptions invoked.
    Where these assumptions are inadequate corresponding approximations should be omitted.}
  \label{fig:approximations}
\end{figure}

BICs and BSCs can be determined by a variety of methods.
We discuss both a direct approach based on survey data and indirect approaches where closely related other metrics are converted into BICs through appropriate approximations (summarised in Fig.~\ref{fig:approximations}).
The latter methods provide more than just useful shortcuts to computing BICs.
They also highlight unifying threads running through various approaches of quantifying biodiversity impacts, and, given the fact that BICs are directly derived from a measure of extinction risk, lend additional scientific support to established tools.

The choice of method/approximation will depend on the spatial resolution required, the accuracy sought and the type and quality of the available data.
When simplified methods yield inconsistent results, it is likely that one of the underlying simplifying assumptions linking them to BICs has been violated in an essential way.
Since these assumptions are spelled out in the derivations below, researchers can seek to refine existing tools where these assumptions are violated to align better with the objective of quantifying impact on extinction risk.

\subsection{Direct determination of BICs and BSCs}
\label{sec:direct-comp-bic}

\subsubsection{From bespoke survey data and global population estimates}
\label{sec:from-survey-data}

Direct computation of BICs and BSCs from survey data is most suitable for small sites with good data availability.
As a pilot test on commercial property, we obtained permission from the owner of a grouse moor estate in Yorkshire, UK, to determine BSCs on their estate.
Grouse moors are managed to provide optimal growth conditions for populations of game birds for recreational shooting.
For an in-depth discussion of this practice, see, e.g.\ \cite{GrouseMoorManagementReviewGroup19:_ReportScottish}.
Many grouse moors are run as businesses, and BSCs might provide opportunities and incentives to create additional value through biodiversity protection.

A commercial biodiversity survey provider (Ecology Services Ltd) was commissioned to conduct bird surveys on a representative area ($1.7\,\mathrm{km}^2$ size) within the estate.
One-hour surveys were conducted on four days in the spring of 2022 (9, 21 April and 13, 26 May).
To conduct the surveys, the area was divided into 5 sections along clearly visible boundaries and consecutively for each section the birds present in it at one moment were counted by observation from an adjacent road.

BSCs were determined for each day, inserting observed population sizes as $n_i$ in the BSC definition, Eq.~\eqref{eq:BSC}. Chicks were included where visible, but did not affect the counts in a relevant way.
For the global population sizes $N_i$ in Eq.~\eqref{eq:BSC} we used the central estimates of \cite{Callaghan21:_GlobalAbundance} \citep[see also][]{Robinson22:_ExtremeUncertainty,Callaghan22:_ReplyRobinson}.
The globally rarest observed species was the stock dove (\textit{Columba oenas}) with an estimated global abundance of 5~million individuals.
For comparison, typical estimates for the regularisation constants $N^*=v_{\mathrm{d}}/v_{\mathrm{e}}$ for birds are in the range of 2.6--60 individuals \citep{Saether04:_LifeHistory, engen09:_reprod_value_and_stoch_demog,Saether13:_HowLife}, i.e., much smaller.
We therefore disregarded the regularisation constants here and, by the same reasoning, in all subsequent calculations.

Averaging over the four days, we calculated a bird BSC density of $1.82 \pm 0.37\times 10^{-6}\mathrm{species}\,\mathrm{km}^{-2}$.
Just 14 species contributed over 99\% to this score -- a patterns reflecting the known high skews of local and global species abundance distributions \citep{Enquist19:_CommonnessRarity, Callaghan21:_GlobalAbundance} and expected to be typical for BSCs.
Here, the metric was dominated by black-headed gull (\textit{Chroicocephalus ridibundus}), which contributed 33\%, willow ptarmigan (\textit{Lagopus lagopus}) contributing 23\%, and Eurasian curlew (\textit{Numenius arquata}) contributing 18\%.
While Eurasian curlew is listed as Near Threatened on the IUCN Red List (REF), the reason is not a low population ($\sim 26$ million) but a rather high rate of population decline.
The next largest contributions came from European golden plover (\textit{Pluvialis apricaria}, 7\%) and stock dove (5\%).
The latter was represented by only 4 of a total of 705 observations (0.6\%) but contributed disproportionately due to its low global abundance.
This finding might indicate a potential to increase BSCs at the site by creating conditions somewhat more suitable for this comparatively rare species.




\subsection{Indirect computation from established metrics}
\label{sec:indirect-comp-bic}

\subsubsection{From Range Size Rarity}
\label{sec:appr-const-popul}

For practical reasons, conservation ecologists often work with presence/absence data of species at lattice elements, disregarding population density.
A biodiversity metric often computed from such data is Range-Size Rarity (RSR) \citep{Howard91:_NatureConservation, Williams94:_CentresSeed-plant}, also known, e.g., as `endemism richness' \citep{Kier09:_GlobalAssessment} and `Rarity Score' \citep{Possingham00:_MathematicalMethods}.

Denote by $C_i$ the index set of the lattice elements occupied by species $i$ and by $S_e$ the set of (the indices of) the species present at lattice element $e$, so that $i \in S_e$ if and only if $e \in C_i$.
We shall write $|\cdot|$ to denote the number of elements of a set.
For example, $|C_i|$ is the number of lattice elements in which species $i$ is present---a measure of range size. RSR at a given lattice element $e$ is then defined as the sum of the inverse range sizes of all extant species \citep{Williams94:_CentresSeed-plant}:
\begin{align}
  \label{eq:RSR}
  \text{RSR}_e = \sum_{\substack{i \in S_e}}\frac{1}{|C_i|}.
\end{align}
RSR has been interpreted as a measure of local species richness where species are weighted by inverse range size, a measure of endemism \citep{Williams94:_CentresSeed-plant}.
\cite{Kier01:_MeasuringMapping} propose further approximations to simplify computation of this metric.

RSR and variants thereof \citep{Guerin15:_SumInverse} are often considered in conservation ecology, especially in the selection of protected areas for species conservation.
In the simplest case, one would place protected areas in the subset of lattice elements with the highest RSR, with the size of this subset depending on conservation effort.

To derive the approximate relation between RSR and BSCs, we assume Earths' surface to be covered by a lattice of $E$ non-overlapping surface elements of equal area $\Delta A$ and, to the degree possible, approximate square shape.
We denote by $n_{i,e}$ the number of individuals of species $i$ in lattice element $e$ and by $\Delta n_{i,e}$ changes in these numbers.
Summing over all lattice elements, $N_i = \sum_e^E n_{i,e}$.

To emulate presence/absence (rather than abundance) data in our analysis, assume that each species $i$ has, wherever it is present, a constant density $\rho_i$.
That is, either $n_{i,e}=\rho_i\Delta A$ or $n_{i,e}=0$.
In this approximation, $N_i=\rho_i|C_i|\Delta A$.

For the following demonstration of the close relation between RSR, BICs and BSCs, we consider, for simplicity, only the case where the approximation of $\text{BIC}_{\alpha}$ as $\Delta \text{BSC}_{\alpha}$, Eq.~\eqref{eq:BIC-vs-BSC}, holds.
In particular, we assume that changes in global abundances are relatively small.
To compute BICs using RSR, Eq.~\eqref{eq:BIC}, we further assume that the site $\alpha$ is given by one or more lattice elements and write $e \in \alpha$ to express that an element $e$ is contained in $\alpha$.
We introduce indicator variables $x_{i,e}$ such that $x_{i,e}=1$ if the impact leads to extirpation of species $i$ in element $e$ and $x_{i,e}=0$ otherwise, so that $\Delta n_{i,e}=-\rho_ix_{i,e}\Delta A$.
Thus $\Delta n_{\alpha,i}=\sum_{e \in \alpha}\Delta n_{i,e}=-\sum_{e \in \alpha}\rho_ix_{i,e}\Delta A$.
Above assumption that changes in the $N_i$ are small implies that most $x_{i,e}$ are zero.
Disregarding the regularisation constants $N_i^*$, we can then write
\begin{align}
  \label{eq:lattice-impact2}
  \text{BIC}_\alpha\approx\sum_i^S \frac{\Delta n_{\alpha,i}}{N_i}\approx -\sum_i^S \frac{\sum_{e \in \alpha}\rho_ix_{i,e}\Delta A}{\rho_i|C_i|\Delta A} = -\sum_{e\in \alpha}\sum_{i}^S\frac{\rho_ix_{i,e}\Delta A}{\rho_i|C_i|\Delta A}=-\sum_{e\in\alpha}\sum_{i}^S\frac{x_{i,e}}{|C_i|}. 
\end{align}
Analogously, 
\begin{align}
  \label{eq:lattice-impact-off}
  \text{BIC}_{\text{off}} \approx -\sum_{e\not \in \alpha}\sum_{i}^S\frac{x_{i,e}}{|C_i|}.  
\end{align}
The crucial observation at this step is that each species' density $\rho_i$ and the size of lattice elements $\Delta A$ cancel out.

As a simple example, consider the case where site $\alpha$ becomes uninhabitable to all species in the group of interest (e.g.\ due to conversion of natural land to intensively farmed land).
All species present at $\alpha$ disappear from that site, implying that $x_{i,e}=1$ for all $e\in \alpha$ and all species $i\in S_e$, and otherwise $x_{i,e}=0$.
Then we get from Eqs.~\eqref{eq:lattice-impact2}, applying Eq.~\eqref{eq:RSR},
\begin{align}
  \label{eq:simple-rarity-impact}
  \text{BIC}_{\alpha} \approx -\sum_{e\in \alpha}\sum_{\substack{i \in S_e}}\frac{1}{|C_i|} = -\sum_{e\in \alpha} \text{RSR}_e.
\end{align}
The simple case of rewilding can be handled analogously.

For more detailed calculations tools such as GLOBIO-Species (\url{https://www.globio.info/}) can be used that can incorporate the effect of land-use type on species ranges \citep{Gallego-Zamorano20:_CombinedEffects} and to some degree also on population densities.

The interpretation of BICs as a predictor of species survival implies that selection of protected areas based on RSR is the strategy that minimises mean long-term species extinction risks if range maps are the only data available.
This result substantiates intuitive arguments conservation ecologists have invoked since \cite{Howard91:_NatureConservation} to justify use of RSR in conservation decisions.

A simple way to understand Eq.~\eqref{eq:simple-rarity-impact} is to notice the analogy between BSCs and RSR: while BSCs are given by the sum of the proportions that the focal area contributes to the global \emph{populations} of species, RSR are defined as the sum of the proportions that the area contributes to the total \emph{ranges} of species.
In the approximation that species are evenly distributed over ranges, the BSCs of a lattice element equal its RSR.

It follows that the global distribution of RSR shown in Fig.~\ref{fig:RSR-STAR} provides a good representation of the global distribution of BSCs.
Considering the logarithmic colour scale used, it becomes clear that globally BSC are concentrated in a few geographically constrained biodiversity hotspots, typically at low latitudes.

\subsubsection{From the STAR metric}
\label{sec:star-as-variant}

\begin{figure}
  \centering
  \includegraphics[width=\linewidth]{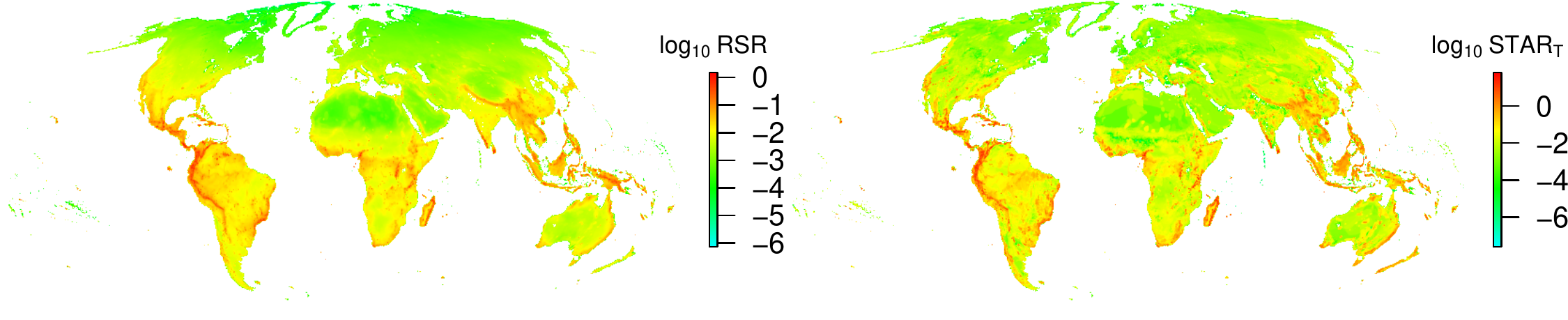}
  \vspace*{-0.5cm}
  \caption{\label{fig:RSR-STAR}  [colour in print] Comparison of Range-Size Rarity (RSR) and Species Threat Abatement and Recovery ($\text{STAR}_{\text{T}}$) metrics for amphibians, birds and mammals.
    Spearman's rank correlation of the two global data sets is $0.80$. Logarithmic major axis regression yields $\text{STAR}_{\text{T},e}\propto (\text{RSR}_{e})^{1.51}$.
    The $\text{STAR}_{\text{T}}$ data was published by  \cite{Mair21:_MetricSpatially}, the RSR data was provided by IUCN (\url{https://www.iucnredlist.org/resources/other-spatial-downloads}). See SI for the detailed method of data analysis and comparison.}
\end{figure}

The Species Threat Abatement and Recovery (STAR) metric  \citep{Mair21:_MetricSpatially} is a weighted variant of RSR. The threat-related STAR (summed over all threats), for example, is given by
\begin{align}
  \label{eq:star-T}
  \text{STAR}_{\mathrm{T},e}=\sum_{\substack{i\\e \in C_i}}\frac{W_i}{|C_i|},
\end{align}
with the weighing factors $W_i$ quantifying the IUCN Red-List category of species $i$ by an integer ranging from $0$ (Least Concern) to $4$ (Critically Endangered).

IUCN's threat-level categories are determined not only based on population size, but also also on other factors affecting the likelihood of future population declines, such as past or current population decline or the fragmentation of populations \citep{Mace08:_QuantificationExtinction}.
Population size itself enters the $\text{STAR}_{\text{T}}$ metric twice. As an important determinant of threat level $W_i$ and in the form of range size $|C_i|$. Indeed, $W_i$ and logarithmic range size are correlated (see Extended Data Figure~4b in  \citealt{Mair21:_MetricSpatially}).
Because $|C_i|$-values vary over many orders of magnitude, most of the variation in $\text{STAR}_{\text{T}}$ is due to $|C_i|$ rather than $W_i$.
As a result, the global patterns of variation in $\text{STAR}_{\text{T}}$ and RSR, and so in BSC, are very similar (Fig.~\ref{fig:RSR-STAR}).

Unfortunately, the published global data sets of RSR and $\text{STAR}_{\text{T}}$ available to us have been generated using incompatible methodologies. We therefore could not determine a valid statistical relationship that would permit numerical conversion between the two metrics, but hope that future work will close this gap.

An immediate implication of the formal similarity of RSR and STAR is that wherever STAR can be computed RSR can be computed to a similar accuracy.
Since STAR assigns weight $0$ to species of Least Concern, data deficiency in this category could in principle lead to additional uncertainty in RSR.
In general, however, these species will have large ranges and so make relatively small contribution to RSR, limiting the resulting uncertainty.
In fact, the IBAT tool (\url{www.ibat-alliance.org}) used to compute STAR can output RSR values as well.

\subsubsection{From Range Size Rarity and the Potentially Disappeared Fraction of species}
\label{sec:appr-rand-extirp}

As explained above, the off-site biodiversity impacts of business activity can be spread out widely across the world.
Even when impacts at any given location are small, the accumulated global impact may be large.
This is the kind of problem often considered in life-cycle assessments.
In this context, the concept of the \emph{Potentially Disappeared Fraction} of species (PDF) has been introduced to quantify diffuse impacts of products on the environment \citep{Muller-Wenk98:_LandUse, Goedkoop00:_Eco-Indicator99}.
The metric is defined as the proportion of locally extant species that get extirpated (i.e., `disappear') as a result of exposure to a pressure such as environmental pollution.
The local `disappearance' of species quantified by PDF is considered reversible once the pressure has ceased.

Denote by $\text{PDF}_e$ the potentially disappeared fraction of species at lattice element $e$.
Now, recall that if a species $i$ locally disappears (i.e., is extirpated), then $x_{i,e}$ in Eq.~\eqref{eq:lattice-impact2} is one, and otherwise it is zero.
We therefore have on average
\begin{align}
  \label{eq:PDF-x}
  \text{PDF}_e = \frac{1}{|S_e|}\sum_{i\in S_e} x_{i,e}.
\end{align}

PDF and RSR together can be used to approximate $\text{BIC}_{\text{off}}$.
This approximation depends on the assumption that disappearance of species at a site where it is present is uncorrelated with its inverse range size, that is, that we can approximate the mean of $x_{i,e}|C_i|^{-1}$ at $e$ over all species $i$ by the product of the means of the two factors.
With $|S_e|$ denoting local species richness, we can then write 
\begin{align}
  \label{eq:uncorrelated}
  \frac{\sum_{i\in S_e}x_{i,e}|C_i|^{-1}}{|S_e|}
  \approx
  \frac{\sum_{i\in S_e}x_{i,e}}{|S_e|}
  \frac{\sum_{i\in S_e}|C_i|^{-1}}{|S_e|}.
\end{align}
This assumption is justified, e.g., when, as recent results suggest \citep{OSullivan19:_Metacommunity-scaleBiodiversity, OSullivan22:_TemporallyRobust}, population and range sizes of species are to a large part controlled by complex ecological interaction networks rather than by the species' own traits, and so effectively random.

Equations~\eqref{eq:uncorrelated},~\eqref{eq:PDF-x} and~\eqref{eq:RSR} allow us to evaluate Eq.~\eqref{eq:lattice-impact-off} as 
\begin{align}
  \label{eq:pdf-impact}
  \text{BIC}_{\text{off}} \approx
  -\sum_{e\not \in \alpha}^E\sum_{i}^S\frac{x_{i,e}}{|C_i|}
  =
  -\sum_{e\not \in \alpha}^E\sum_{i\in S_e}\frac{x_{i,e}}{|C_i|}
  \approx
  -\sum_{e\not \in \alpha}^E\frac{\sum_{i\in S_e}x_{i,e}}{|S_e|} \sum_{i\in S_e}\frac{1}{|C_i|}
  \approx
  -\sum_{e\not \in \alpha}^E\text{PDF}_e\, \text{RSR}_e.
\end{align}


\subsubsection{From species density and PDF}
\label{sec:from-species-density}

RSR is closely related to species density \citep{Kier01:_MeasuringMapping}.
To see this, we compute the sum of $\text{RSR}_e$ over a hypothetical large area in which all species are endemic.
We denote the set of lattice elements forming this area by $\Omega$ and the set of endemic species by $S_\Omega$.
Then
\begin{align}
  \label{eq:RSR-SD}
  \sum_{e\in\Omega} \text{RSR}_e=\sum_{e\in\Omega} \sum_{i\in S_e}\frac{1}{|C_i|} =
  \sum_{i \in S_\Omega}\sum_{e\in C_i} \frac{1}{|C_i|}=
  \sum_{i \in S_\Omega} \frac{1}{|C_i|}\sum_{e\in C_i}1=\sum_{i \in S_\Omega} 1=|S_\Omega|.
\end{align}
Dividing both the first and the last expression by the size of the area $A=|\Omega|\Delta A$, we see that species density $|S_\Omega|/A$ equals the average of $\text{RSR}_e$ over $\Omega$ in units of $\Delta A$.
In particular, the global sum of RSR equals global species richness $S$.
Even when the ranges of a few species extend beyond the area considered, Eq.~\eqref{eq:RSR-SD} holds to a good approximation.
Similar arguments apply to BSCs.
On large scales, BSC density, RSR/$\Delta A$, and species density therefore become largely indistinguishable.

For pollutants that spread globally before they decay, we may assume that $\text{PDF}_e$ is constant or at least uncorrelated with $\text{RSR}_e$.
Disregarding the exclusion of the (usually small) site $\alpha$ in Eq.~\eqref{eq:pdf-impact}, we can then replace $\text{PDF}_e$ by its global average $\overline{\text{PDF}}$ to obtain, using Eq.~\eqref{eq:RSR-SD},
\begin{align}
  \label{eq:pdf-impact2}
  \text{BIC}_{\text{off}} \approx
  -\sum_{e}^E\text{PDF}_e\, \text{RSR}_e\approx
  -\sum_{e}^E \overline{\text{PDF}}\,\text{RSR}_e=
  -\overline{\text{PDF}} \sum_{e}^E \,\text{RSR}_e=
  -S\,\overline{\text{PDF}}.
\end{align}
When $\text{PDF}_e$ and species density are correlated, e.g., on continental scale, expressions such as the right-hand-side of Eq.~\eqref{eq:pdf-impact2} should be evaluated for each continent separately and then added up.

An alternative formulation of Eq.~\eqref{eq:pdf-impact2} expresses species richness $S$ in terms of global species density $\text{SD}=S/(E\Delta A)$, recalling that $E$ is the total number of lattice elements covering Earth's surface:
\begin{align}
  \label{eq:pdf-impact3}
  \text{BIC}_{\text{off}} \approx
  -S\,\overline{\text{PDF}}=
  -\text{SD} \,E \,\Delta A \,\frac{\sum_e^E\text{PDF}_e}{E}=
  -\text{SD} \,\sum_e^E\text{PDF}_e \,\Delta A.
\end{align}
In practical calculations care must be taken to average and sum only over land surfaces, only over water surfaces, or consistently over both.

\subsubsection{Relations to the work of \cite{Kuipers19:_PotentialConsequences} and \cite{Verones22:_GlobalExtinction}}
\label{sec:relation-work-}

It is worth noting the similarity between the sum on the right-hand side of Eq.~\eqref{eq:pdf-impact} and Equation~(2) of \cite{Verones22:_GlobalExtinction}, which formalises a corresponding verbal prescription by \cite{Kuipers19:_PotentialConsequences}.
Both differ in three rather minor points from our Eq.~\eqref{eq:pdf-impact}: (i) in place of $\text{RSR}$ an early variant of $\text{STAR}_{\text{T}}$ containing weighting by Red-List Categories is used, (ii) this variant metric (called `Global Extinction Probability', although it was not derived as a probability) is normalised such that its sum over the globe is 1, and (iii) rather than summing over lattice elements, the sum is over a partition of Earth's land surface into ecological coherent regions (ecoregions).
\cite{Kuipers19:_PotentialConsequences} interpreted this sum as ``translating fractions of regional species extinctions into global species extinctions'', arguing for this interpretation by considering situations where all species are lost from one or several ecoregions (i.e., $\text{PDF}=1$ in these regions).

This interpretation is correct without the weighting by Red-List Categories (i.e., when using $\text{RSR}$) and when these ecoregions contain only endemic species (in which case $\text{RSR}$ reduces to regional species richness).
For the more common situation with non-endemic species and PDF values between 0 and 1 (which then refer to local rather than regional extirpations), our interpretation of Eq.~\eqref{eq:pdf-impact} as a measure of impact on the risk of future extinctions (rather than a measure of current extinctions) appears appropriate in view of its mathematical derivation above.
This refined interpretation notwithstanding, the work by \cite{Kuipers19:_PotentialConsequences} and \cite{Verones22:_GlobalExtinction} has demonstrated the strong intuitive appeal of Eq.~\eqref{eq:pdf-impact} and related results presented above.

\subsubsection{From PDF-based metrics in species$\times$year units}
\label{sec:appr-range-size}

PDF-based footprinting tools such as \texttt{ReCiPe 2016} \citep{Huijbregts17:_ReCiPe2016Harmonised}, which have their roots in life-cycle impact assessment, allow users to compute a measure of the ecosystem damage ($\text{ED}$) resulting from a single unit of business activity.
Such a unit may be the production, sale and use of a single product, emission of one unit of pollutant, or one year of business activity.
Since the environmental impact of one such unit will eventually, if slowly, decline as pollutants decay and mineralise and ecosystems recover, ecosystem damage is not measured as static global mean PDF value but in terms of the integral of a dynamically declining impacts over time.
As a result, units of impact generally contain a factor `year'.

$\text{ED}$, for example, is computed by a formula analogous to the right-hand-side of Eq.~\eqref{eq:pdf-impact3}, where in place of $\text{PDF}_e$ one finds a sum of impacts along different pathways (e.g.\ different pollutants) in units of $\text{PDF}\times\text{year}$.
Hence, while the right-hand-side of Eq.~\eqref{eq:pdf-impact3} has units of species, $\text{ED}$ is expressed in units of species$\times$year.

The simplest way to convert time-integrated metrics into metrics of static impact is to multiply them with the rate $r$ at which units of business activity are produced ($r=1\,\mathrm{year}^{-1}$ when the unit is all activity over one year), which yields the long-term impact that would result if the business activity would progress indefinitely at a constant rate.
We can then approximate $\text{BIC}_{\text{off}}$ in terms of $\text{ED}$ as
\begin{align}
  \label{eq:ed-to-bic}
  \text{BIC}_{\text{off}}\approx -\text{SD} \,\sum_e^E\text{PDF}_e \,\Delta A=  \text{ED}\,r.
\end{align}
More detailed calculations would take the time scales of accumulation and decline of various pressures on biodiversity into account, e.g., according to the models underlying \texttt{ReCiPe 2016}.

\section{Discussion}
\label{sec:discussion}

\subsection{A strictly science-based target}
\label{sec:strictly-science-bas}

The concept of a science-based target \citep{Andersen21:_DefiningSciencebased} has three distinct aspects: it is achievable, there is a process permitting a scientifically meaningful answer to the question whether it has been met, and ``a clear, analytical rationale'' supports the target level.
For discussion of the process we refer to \cite{deSilva19:_EvolutionCorporate}.
Target setting will often build on the principle of intergenerational justice \citep{Rossberg17:_QuantitativeCriteria}.
In the context of climate-change mitigation, the general principle \citep{Bjorn21:_ParisAgreement} is to set a limit to global warming (protecting future generations) and to allocate the corresponding remaining greenhouse-gas emission budget over current emitters (benefiting current generations).

The \cite{ScienceBasedTargetsNetwork23:_ResourceRepository} has the ambition to set impact targets for species conservation following a similar logic.
The current draft Technical Annex underlying their methodology (published in a version dated September 2020 on their website), however, acknowledges a prevailing knowledge ``gap'' that prevents this ambition from being realised.
While the Annex discusses workarounds based in established conservation ecology, establishment of science-based targets for species conservation remains a challenge.

It is the absence of such a gap that distinguishes BICs and the associated target, Eq.~\eqref{eq:offsetting-equation} from comparable metrics.
BICs are linked mathematically through Eq.~\eqref{eq:offsetting-theorem} to changes in $\mathcal{L}_{\text{reg}}$, the quantity $\mathcal{L}_{\text{reg}}$ is linked in Box~\ref{box:regularisation} to a mathematical model for extinction risk, and the model builds on the empirically well-documented phenomena of environmental and demographic stochasticity.
This rigour motivates our characterisation of BICs and the associated target as being `strictly science-based'.
It supports, furthermore, established intuitive motivations for several metrics quantitatively linked to BICs (Fig.~\ref{fig:approximations}) with a stronger rationale.

\subsection{BICs in a nutshell}
\label{sec:bic-nutshell}

The use of BICs by business is based on two simple messages: (1) When $\text{BIC}_{\alpha}+\text{BIC}_{\text{off}}>0$ the mean long-term survival probability of species increases as a result of the business's activities.
Businesses that are biodiversity positive in the sense that $\text{BIC}_{\alpha}+\text{BIC}_{\text{off}}>0$ contribute to overcoming the global extinction crises rather than being part of the problem.
(2) The value of $\text{LPI}$ (as fundamentally defined; not necessarily empirical estimates from population time series) increases as a result of the business's activities by a proportion
\begin{align}
  \label{eq:LPI-form-BIC}
  \frac{\Delta \text{LPI}}{\text{LPI}}\approx\frac{\text{BIC}_{\alpha}+\text{BIC}_{\text{off}}}{S},
\end{align}
where $S$ is the total number of species in the group considered.
Biodiversity positive businesses in the sense above therefore increase the value of this widely cited measure of biodiversity loss.

\subsection{Offsetting or not?}
\label{sec:offsetting-or-not}

Businesses have a wide range of options for positioning themselves in the debate on how exactly to achieve `nature positive' outcomes \citep{zuErmgassen22:_AreCorporate}.
These include: achieving positive on-site biodiversity impact on their own or after adding bought-in credits; achieving this objective after inclusion of off-site impacts; precautionary approaches that apply a multiplier to negative impacts before adding the positives \citep{Bull20:_NetPositive}; reporting on negative and positive impacts without computing a net impact; emphasising systemic change to address externalities \citep{ScienceBasedTargetsNetwork21:_NaturepositiveOpportunity}; or decoupling the discourse on positive impacts entirely from the consideration of negatives \citep{WorldEconomicForum22:_BiodiversityCredits}. 
The metrics proposed here are applicable in any of these contexts and we do not advocate for any particular position.

However, a clarification of terminology is in place.
The term `biodiversity offset' tends to be associated with traditional schemes focused on on-site impacts, emphasising ecosystem function and services and like-for-like compensation \citep{Bull13:_BiodiversityOffsets}.
Zu Ermgassen et al. (\citeyear{zuErmgassen22:_AreCorporate}) argued for moving away from such schemes, based on the recognition that ``footprinting assessments across companies' entire value chains are showing that direct operational impacts are a relatively small part of total biodiversity impacts for many large organisations''.
The objective of 'nature-positive' outcomes should instead draw attention to wider impacts and societal objectives.
We note, though, that the `net' in any conceptualisation of `Nature Positive' involving ``net gain in biodiversity'' \citep{Milner-Gulland22:_DonDilute} implies some form of arithmetic combining positives and negatives, whether one calls this `offsetting' or not.
Rather than attempting to coin new terms, we here follow \cite{Bull22:_AnalysisBiodiversity} in sticking to the term `offset' even if applied in the context of the Nature Positive agenda.
We argue below that, this terminology notwithstanding, the offsetting of BICs can overcome several of the issues associated with traditional offsetting schemes.
This might encourage businesses to quantitatively compare their positive and negative impacts, provide materiality to biodiversity credits \citep{WorldEconomicForum22:_BiodiversityCredits}, and improve resource allocation to species conservation by stimulating entrepreneurial thinking to find cost efficient ways of achieving this in the context of other needs such as ecosystem services.

\subsection{Choice of groups of species to cover}
\label{sec:choice-group-species}

A question we did hitherto not address is that of the choice of the group of $S$ species entering an implementation of BICs or BSCs.
For established related metrics such as PDF-based metrics, LPI, or STAR, this choice appears to have been made based on the availability of suitable data. Since data requirements differ between metrics, so do the taxa or functional groups they cover.

We will not recommend a specific choice for BICs and BSCs at this stage.
This choice should consider amongst others (i) availability of data or technology to determine the metrics or approximate variants, (ii) alignment with relevant established variants, (iii) inclusion of vulnerable taxa, such as Cycads or Amphibians.

One might consider computing and trading BICs separately for two or more distinct groups of species, especially when there are concerns that some groups require more protection than others.
On the other hand, businesses are unlikely to be able to consider biodiversity constraints imposed by very many metrics, and best results might be achieved if BICs are reported and traded for a single but encompassing species group.

Relevant in this context is that RSR and species richness are known to be correlated across taxonomic and functional groups, especially on larger scales and when these groups are not too narrowly defined \citep{Warren92:_Predator-preyRatios, Kier09:_GlobalAssessment, Qian10:_SpatialScale, Castagneyrol12:_UnravelingPlant-animal}.
As this correlation is causally explained by interactions between groups \citep{Castagneyrol12:_UnravelingPlant-animal, rossberg13:_food_webs_biodiv, Zhang18:_TrophicInteractions} there can be a degree of equivalence of metrics covering different groups of species.
The similarity of RSR and BSC suggests that, to some extent, conversion between BIC metrics defined over different groups might also be possible.


\subsection{A worked example}
\label{sec:worked-example}

Let's consider an example. In a recent self-assessment using the \texttt{ReCiPe 2016} tool, University of Oxford reported to be generating an ecosystem damage ($\text{ED}$) of $1.6\,\mathrm{species}\times \mathrm{year}$ in the academic year 2019-20 \citep{Bull22:_AnalysisBiodiversity}.
As this quantifies the summed impact for one year of activity, we can estimate the corresponding steady-state impact by setting $r=\mathrm{year}^{-1}$ in Eq.~\eqref{eq:ed-to-bic}, giving an estimated offsite impact of $\text{BIC}_{\text{off}}\approx -1.6\,\mathrm{species}$.

To compensate this increase in species extinction risk, University of Oxford could acquire BICs from a conservation NGO such as the Tree Conservation Fund.
With funding from University of Oxford, the Fund and its partners would, say, acquire an area of land suitable for the reintroduction of Chinese watermelon trees (\emph{Artocarpus nanchuanensis}), which are critically endangered with perhaps only 100 individuals remaining in the wild.
They might plant 1000 saplings of \emph{A.\ nanchuanensis} in this area.
Assume that when the trees mature after about five years over 200 saplings survive, increasing the tree's mature population on the land held by the fund or its partners by $\Delta n=200$ and bringing the total global mature population to about $N=300$.
Assume further a regularisation constant of $N^*\approx 100$ mature individuals for this species.
The resulting BICs would then evaluate by Eq.~\eqref{eq:BIC} to
\begin{align}
  \label{eq:nanchuanensis}
  \text{BIC} \approx \frac{\Delta n}{N^*+N} \approx \frac{200}{100 + 300} = 0.5\,\mathrm{species},
\end{align}
where we disregarded impacts on other species, as these would likely be dwarfed by the \emph{A.\ nanchuanensis} contribution (except for species dependent on \emph{A.\ nanchuanensis}).
The Tree Conservation Fund can provide this contribution to species conservation at an estimated cost of just around USD~250,000 \citep{TreeConservationFund21:_ArtocarpusNanchuanensis}.

Supporting four projects of this kind, University of Oxford might be able to obtain $\mathrm{BSC}_\alpha=2\,\mathrm{species}$ at a cost of around USD~1,000,000 after five years, thus having turned a negative into a positive impact on the long-term survival of species, quantified by BICs as $\text{BIC}_{\text{off}}+\mathrm{BSC}_\alpha\approx 0.4\,\mathrm{species}$, with some margin of error.

The range of taxa considered by \texttt{ReCiPe 2016} is very broad, including insects, plants, arachnids and vertebrates amongst others and covering $S=1.85\times 10^6$ known species \citep{Goedkoop13:_Recipe2008}.
For an LPI computed for this large group the resulting proportional increase is,  by Eq.~\eqref{eq:LPI-form-BIC}, $0.4\,\mathrm{species}/ S\approx 2\times 10^{-7}$.

\subsection{A dynamic economy of biodiversity}
\label{sec:from-planned-dynamic}


To meet the needs of the business community, the BIC metric has deliberately been designed to be simple and widely applicable and to rely as little as possible on expert input.
Rather than incorporating the complexities and uncertainties of ecological and social systems in the metric's definition, the decision on how best to deal with these is intentionally left to the users of the metric.

There will, for example, always be some uncertainty as to whether restoration measures will generate anticipated gains in BICs. Rather than awarding credits to plans to restore populations and ecosystems and the resulting projected gains, BICs and BSCs are strictly linked to the documented current state.
This does not stop business partners from agreeing on restoration plans and making commitments for gains in BIC, the associated uncertainty is just not part of the metric's definition.
Instead, allocation of risks and opportunities amongst business partners should be part of contractual agreements.
To stimulate entrepreneurial spirit, an agreement might, for example, permit an organisation growing BICs for a client to sell BICs grown beyond the contractual agreement to third parties, while also leaving it responsible to compensate the client for underachievement.
Alternatively, BIC growers might act as service providers, leaving any risks to the commissioning clients, while the latter might hedge these risks by investing into a diverse portfolio of BIC-growing projects.

This flexibility arises because the BIC metric measures a wide variety of positive and negative impacts in the same, convertible currency, thus providing the foundations of a wide BIC market capable of absorbing unexpected gains and losses.
Critics might argue that this flexibility constitutes a weakness of the metric: how can rebuilding one species' population in Malaysia compensate the decline of another species' population in Guatemala?
Doesn't the metric's proposed use as a commodity imply that species are substitutable with each other, while in fact the irreplaceability of each species provides the moral underpinning of biodiversity conservation at the species level?

Such arguments, however, are likely due to misunderstandings.
A first misunderstanding might be that our scheme aims fundamentally at the maintenance of species populations.
This is not the case.
The scheme aims to reduce the risk of species extinctions; keeping species populations large is just a means to this end.

A second misunderstanding might be that BICs directly trade off gains in one population against losses in another.
This would indeed be inadequate for species conservation, especially if the species experiencing losses is the rarer one of the two. In fact, however, rarer species enter BIC with a higher weight, which intrinsically encourages rebuilding their populations.

Finally, should the above critical argument relate to a situation where both the Malaysian and the Guatemalan species are rare and at similarly high risk of extinction, its fallacy lies in the assumption that sufficient funding to protect these two species and all other species with similar status is available.
Extinction rates are high \citep{Ceballos15:_AcceleratedModern} and global resources for conservation insufficient and inadequately allocated \citep{Deutz20:_FinancingNature, Rodriguez22:_AddressingBiodiversity}.
A choice of how much to invest into each species is required.
We invoked a substitutability argument at this point only insofar as we assumed the extinction of one species to be just as undesirable as that of any other (within the group considered).
Resources should therefore be allocated such as to maximise overall expected long-term species survival, even if this means leaving some species at a high risk of extinction.


Intuitively, above example of rebuilding the \emph{A.\ nanchuanensis} population might strike readers as somewhat artificial, as it barely generates any of the benefits we are used to associate with the restoration of nature.
Readers may also wonder how a tree restoration project in China can compensate for damage to the environment done by an institution in England.
We believe that such concerns arise from confounding species conservation with the conservation of ecosystems and ecosystem services.
Indeed, service provision tends to require complex ecosystems and is often not substitutable over large geographic distances.
For species conservation, neither must be an immediate constraint.
Yet, linkages between BICs, complex ecosystems, and ecosystem services naturally arise.

For species conservation it currently makes sense to focus investments on those species at highest risk of extinction.
BICs provide the market signal to do this where feasible.
Coming back to our example: considering the low RSR and BSC density of England (Fig.~\ref{fig:RSR-STAR}), investments by University of Oxford into species protection abroad can reduce overall extinction risk much more than similar investments to restore habitats in England would.
However, after these low-hanging fruit of species conservation have been picked, more intricate schemes where entire ecosystems are rebuilt are likely to become attractive for BIC generation.
This shift in species conservation efforts from populations to ecosystems would be reinforced by the property of BICs, discussed above, of disincentivising dominant holdings for population gains.

Linkages to ecosystem services arise when organisations develop ``nature-based solutions'' that combine generation of BICs with the provision of ecosystem services (e.g. food, clean water, tourism, carbon sequestration, etc.).
The market price of BICs and the expected benefits from ecosystem service provision (which depend on local conditions) will signal to these organisations how best to allocate their resources and where the highest potential for effective synergies might lie.
Thus, our focus on species protection in the design of BICs does not deny the importance of ecosystem services.
We just consider that their clean conceptual separation is conducive to the efficient conservation of both.

By intentionally addressing species conservation only, the approach we propose differs from most traditional biodiversity offsetting schemes and overcomes some of their potential weaknesses, reviewed, e.g., by \cite{Bull13:_BiodiversityOffsets}.
Our scheme does not involve like-for-like offsetting and so avoids the question to which extent this is achieved \citep{Madsen10:_StateBiodiversity}.
This generates a larger market for potential offsets, reducing transaction costs and leading to more efficient resource allocation.
In particular, we expect availability of offsets from biodiversity gains that have already been achieved \citep{Bekessy10:_BiodiversityBank}, thus sidestepping questions related to risk of failure and discount rates.

Two important recommendations, however, apply to our scheme just as much as to traditional offsetting.
The first is that offsets should last at least as long as the negative impacts they are meant to compensate \citep{Bull13:_BiodiversityOffsets}.
Businesses practising offsetting should be ready to demonstrate at any time that they hold sufficient positive BICs to compensate their negative impacts, even when the BICs they hold change due to changes in global species abundances $N_i$.

The second is to maintain reversibility of impacts \citep{Godden03:_TheoreticalIssues}.
On paper it is possible to compensate, in terms of BIC, the deliberate global extinction of one species by rebuilding the populations of several other species close to extinction, and indeed this might increase overall long-term species survival.
Yet, we caution against this based on arguments similar to those invoked in opposition to the death penalty: the irreversibly of the intentional extermination makes it impossible to correct methodological, conceptual, theoretical and implementation errors that may become clear after the scheme's execution.



\bigskip

\section*{Conclusion}
\label{sec:conclusion}

We have demonstrated that BICs have properties suitable for use in biodiversity-related disclosures in business and financial contexts.
The metric can also support voluntary or legislated nature positive policies.
In all cases, we recommend its use in conjunction with metrics of ecosystem extent and integrity.
Many of the attractive properties of BICs reflect that this metric is strictly science based, mathematically linked to the LPI and the species conservation objective.
Data requirements of BICs are similar to those of existing comparable metrics.
Pilot studies are now underway to test marketing of BICs in practice.

\subsection*{Acknowledgements}
\label{sec:acknowledgements}

The authors thank J. Christopher (Chris) D. Terry, Paul Smith, Aafke Schipper, and J.\ P.\ (Jelle) Hilbers for comments on earlier drafts of this paper.
Research supported by the UK's Natural Environment Research Council (grants NE/T003510/1, NE/W00965X/1, NE/X016439/1).

\subsection*{Author Contributions}
\label{sec:author-contributions}

\textbf{Axel G. Rossberg:} Conceptualization, Methodology, Formal analysis, Writing - Original Draft, Review \& Editing, Visualization, Funding acquisition, \textbf{Jacob D. O’Sullivan:} Methodology, Software, Writing - Review \& Editing, Visualization, \textbf{Svetlana Malysheva:} Formal analysis, Data Curation , \textbf{Nadav M. Shnerb:} Methodology, Formal analysis, Writing - Review \& Editing.

\medskip

\noindent The funders had no involvement: in study design; in the collection, analysis and interpretation of data; in the writing of the report; and in the decision to submit the article for publication.



\clearpage

\appendix
\beginappendix

\begin{center}
  \Large Appendices
\end{center}

\section{Numerical test of our model for extinction risk}
\label{sec:numer-test-equat}

Here we demonstrate that Eq.~\eqref{eq:P-surv} provides a good estimate of mean species lifetime despite making use of the diffusion (continuum) approximation and disregarding the decline of the strength of fluctuations near $x=0$.

For this we simulated an asexual population with non-overlapping generations as
\begin{subequations}
  \label{eq:pop-model}
\begin{align}
    \lambda(t) &\sim \exp\!\left(v_{\text{e}}^{1/2}\,\xi(t)\right),\label{eq:lognormal}\\
    N(t+1) &\sim \mathop{\mathrm{Poisson}}\!\left(\lambda(t)\,N(t)\right),\label{eq:poisson}
\end{align}
\end{subequations}
with $\xi(t)$ denoting a standard normal random number and $\mathop{\mathrm{Poisson}}(\mu)$ a Poisson-distributed random number with mean $\mu$.
The model implies $v_{\text{d}}=1$.

We set $v_{\text{e}}=(0.15)^2$ and evaluated 26 values for the initial population size $N(0)=N_0$, spaced equally on the log axis from $1$ to $10^5$, except for rounding to the nearest integer. 
For each $N_0$, the probability of species survival ($N(T)>0$) until time $T=10^4$ was estimated from $10^4$ replicates.

\begin{figure}[b]
  \centering
  \includegraphics[width=0.5\linewidth]{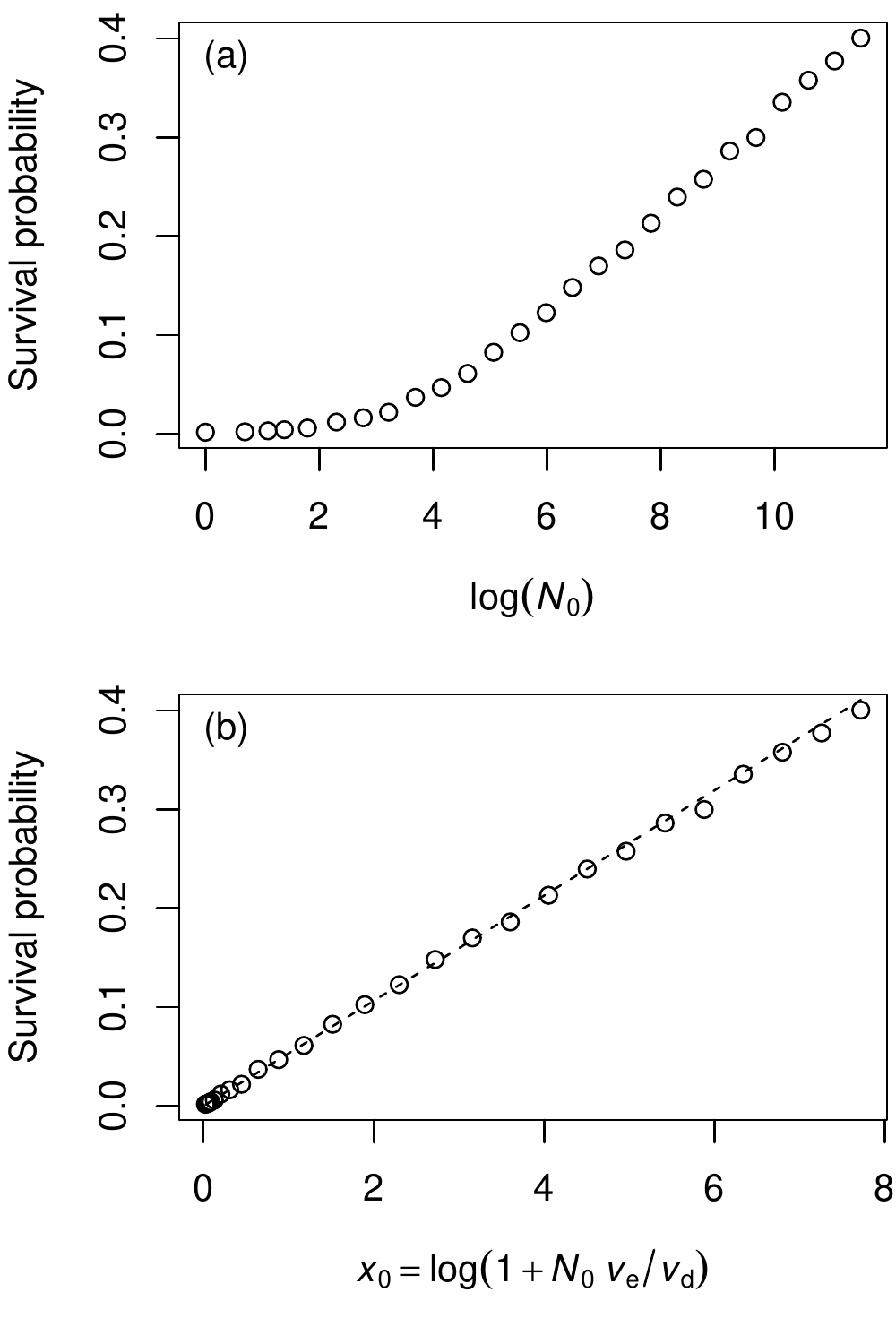}
  \caption{\textbf{Demonstration that $x_0$ predicts expected species lifetime better than $\log(N_0)$.\label{fig:pExt}}  [no colour in print] Points are simulation results for the individual-based model Eq.~\eqref{eq:pop-model} with $v_{\text{e}}=(0.15)^2$ over $T=10^4$ time steps. The dashed line is the probability of survival as approximated by Eq.~\eqref{eq:P-surv}. Panels (a) and (b) represent the same data on different horizontal scales.}
\end{figure}

As shown in Fig.~\ref{fig:pExt}a, extinction probability increases linearly with $\log(N_0)$ only for large $N_0$.
By contrast, $x_0=\log(1+N_0/N^*)$, with $N^*=v_{\text{d}}/v_{\text{e}}=44.4$, is linearly related to extinction probability for all $N_0$ considered (Fig.~\ref{fig:pExt}b; since probabilities are $\le 1$ by definition, the relationship would break down for even larger $N_0$).
The slope of the relation is as predicted by Eq.~\eqref{eq:P-surv}.

\section{Non-additivity of increments $\Delta \mathcal{L}_{\text{reg}}$}
\label{sec:non-additv-chang}

To enable proper trade in biodiversity credits, the underlying metric should be additive in the following sense: Assume an initial situation where market participant A holds $C_{\alpha}$ credits (by some metric) and B holds $C_{\beta}$ credits (possibly, but not necessarily in adjacent sites), for simplicity both computed against the same baseline year.
Now, assume that A buys the site held by B and with it the associated credits.
Denote the combined site by $\gamma$.
The formula to compute biodiversity credits is additive when the credits $C_\gamma$ now held by A according to this formula equal $C_{\alpha}+C_{\beta}$.

BICs have this property.
With $\text{BIC}_\alpha=\sum_i^S \Delta n_{\alpha,i}/(N_i^*+N_i)$ and $\text{BIC}_\beta=\sum_i^S \Delta n_{\beta,i}/(N_i^*+N_i)$, and considering that abundance increments for the combined site are given by  $\Delta n_{\gamma,i}=\Delta n_{\alpha,i}+\Delta n_{\beta,i}$, it follows immediately that
\begin{align}
  \label{eq:addingBICs}
  \text{BIC}_\alpha+\text{BIC}_\beta= \sum_i^S \frac{\Delta n_{\alpha,i}+\Delta n_{\beta,i}}{N_i^*+N_i}= \sum_i^S  \frac{\Delta
  n_{\gamma,i}}{N_i^*+N_i}=\text{BIC}_\gamma.  
\end{align}

We will now show that changes in $\mathcal{L}_{\text{reg}}$ as given by Eq.~\eqref{eq:DeltaL} are not additive in this sense.
Using subscripts $\alpha$, $\beta$, $\gamma$ in the obvious way, we show that $(\Delta \mathcal{L}_{\text{reg}})_\alpha+(\Delta \mathcal{L}_{\text{reg}})_\beta-(\Delta \mathcal{L}_{\text{reg}})_\gamma= 0$ only for particular choices of $\Delta n_{\alpha,i}$ and $\Delta n_{\beta,i}$.

Using Eq.~\eqref{eq:DeltaL}, we evaluate
\begin{multline}
  \label{eq:DeltaLcalc}
  (\Delta \mathcal{L}_{\text{reg}})_\alpha+(\Delta \mathcal{L}_{\text{reg}})_\beta-(\Delta \mathcal{L}_{\text{reg}})_\gamma=\\
  \sum_{i}^S\log\left[\frac{N_i^*+N_i}{N_i^*+N_i-\Delta n_{\alpha,i}}\right] +\sum_{i}^S\log\left[\frac{N_i^*+N_i}{N_i^*+N_i-\Delta n_{\beta,i}}\right] -\sum_{i}^S\log\left[\frac{N_i^*+N_i}{N_i^*+N_i-\Delta n_{\gamma,i}}\right]=\\
  \sum_{i}^S\log\left[\frac{(N_i^*+N_i)(N_i^*+N_i-\Delta n_{\gamma,i})}{(N_i^*+N_i-\Delta n_{\alpha,i})(N_i^*+N_i-\Delta n_{\beta,i})}\right]= 
  \sum_{i}^S\log\left[\frac{(N_i^*+N_i)(N_i^*+N_i-\Delta n_{\alpha,i}-\Delta n_{\beta,i})}{(N_i^*+N_i-\Delta n_{\alpha,i})(N_i^*+N_i-\Delta n_{\beta,i})}\right].
\end{multline}
For the expression on the right-hand-side of Eq.~\eqref{eq:DeltaLcalc} to be zero for general $\Delta n_{\alpha,i}$ and $\Delta n_{\beta,i}$, it must be zero for each element of the sum over $i$.
Since further $\log(x)=0$ implies $x=1$, this leads to a condition
\begin{align}
  \label{eq:da-db-cond}
  \frac{(N_i^*+N_i)(N_i^*+N_i-\Delta n_{\alpha,i}-\Delta n_{\beta,i})}{(N_i^*+N_i-\Delta n_{\alpha,i})(N_i^*+N_i-\Delta n_{\beta,i})}=1
\end{align}
for all $i$. Using elementary algebra, this condition simplifies to
\begin{align}
  \label{eq:da-db-cond2}
  \Delta n_{\alpha,i}\Delta n_{\beta,i}=0.
\end{align}
Since Eq.~\eqref{eq:da-db-cond2} is not satisfied for general $\Delta n_{\alpha,i}$ and $\Delta n_{\beta,i}$, increments  $\Delta \mathcal{L}_{\text{reg}}$ are not additive, as noted in the main text.

\clearpage

\paragraph{Code availability statement}

For review purposes, the R code underlying Appendix \ref{sec:numer-test-equat} has been deposited at \url{http://axel.rossberg.net/pExt.R}. The LVMCM simulation model used to elucidate some properties of the BIC/BSC is available at \url{http://www.github.com/jacobosullivan/LVMCM_src}.




\end{document}